\documentclass[10pt,twocolumn,letterpaper]{article}

\usepackage{iccv}
\usepackage{times}
\usepackage{epsfig}
\usepackage{graphicx}
\usepackage{amsmath}
\usepackage{amssymb}

\usepackage{color}
\makeatletter
\@namedef{ver@everyshi.sty}{}
\makeatother
\usepackage{tikz}
\usetikzlibrary{plotmarks,shapes,calc,positioning,shadows.blur,decorations.pathreplacing,shapes.misc}
\usepackage{etoolbox}
\usepackage{pgfplots}
\usepackage{pgfkeys,pgffor}
\pgfplotsset{compat=1.14}
\usepackage{subfig}
\usepackage{booktabs}
\usepackage{multirow}

\usepackage[pagebackref=true,breaklinks=true,colorlinks,bookmarks=false]{hyperref}

\iccvfinalcopy 

\def\iccvPaperID{17} 

\begin{document}

\title{Breast Tumor Cellularity Assessment using Deep Neural Networks}

\author{Alexander Rakhlin\\
Neuromation OU\\
Tallinn, 10111 Estonia\\
{\tt\small rakhlin@neuromation.io}
\and
Aleksei Tiulpin\\
University of Oulu\\
Oulu 90220, Finland\\
{\tt\small aleksei.tiulpin@oulu.fi}
\and
Alexey A. Shvets\\
MIT\\
Boston, MA 02142, USA\\
{\tt\small shvets@mit.edu}
\and
Alexandr A. Kalinin\\
University of Michigan\\
Ann Arbor, MI 48109, USA\\
{\tt\small akalinin@umich.edu}
\and
Vladimir I. Iglovikov\\
ODS.ai\\
San Francisco, CA 94107, USA\\
{\tt\small iglovikov@gmail.com}
\and
Sergey Nikolenko\\
Neuromation OU, Estonia,\\
Steklov Mathematical Institute \\ at St. Petersburg, Russia\\
{\tt\small sergey@logic.pdmi.ras.ru}
}

\maketitle
\ificcvfinal\thispagestyle{empty}\fi

\begin{abstract}
    Breast cancer is one of the main causes of death worldwide. Histopathological cellularity assessment of residual tumors in post-surgical tissues is used to analyze a tumor's response to a therapy. Correct cellularity assessment increases the chances of getting an appropriate treatment and facilitates the patient's survival. In current clinical practice, tumor cellularity is manually estimated by pathologists; this process is tedious and prone to errors or low agreement rates between assessors. In this work, we evaluated three strong novel Deep Learning-based approaches for automatic assessment of tumor cellularity from post-treated breast surgical specimens stained with hematoxylin and eosin. We validated the proposed methods on the BreastPathQ SPIE challenge dataset that consisted of 2395 image patches selected from whole slide images acquired from 64 patients. Compared to expert pathologist scoring, our best performing method yielded the Cohen\rq
    s kappa coefficient of $0.69$ (vs. $0.42$ previously known in literature) and the intra-class correlation coefficient of $0.89$ (vs. $0.83$). Our results suggest that Deep Learning-based methods have a significant potential to alleviate the burden on pathologists, enhance the diagnostic workflow, and, thereby, facilitate better clinical outcomes in breast cancer treatment.
\end{abstract}

\section{Introduction} \label{introduction}

\begin{figure*}
	\includegraphics[width=\linewidth]{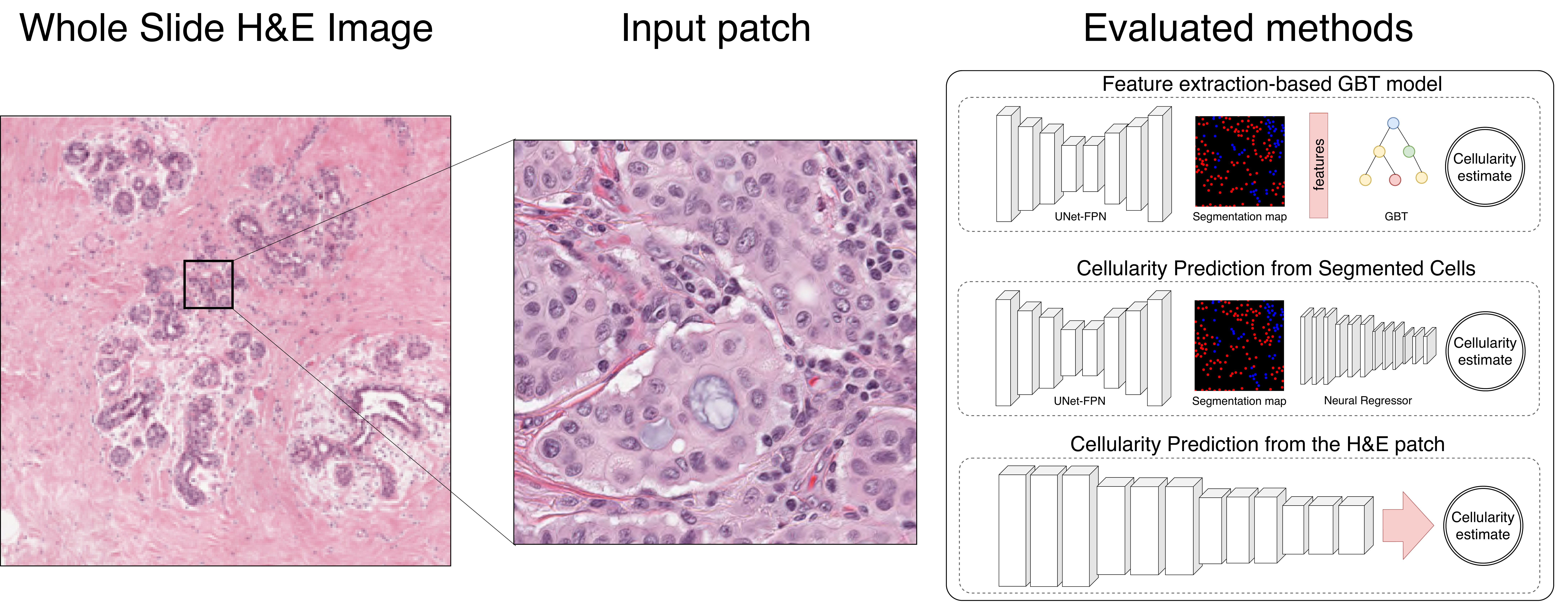}
    \captionof{figure}{Generic description of the methods developed and evaluated in this study. Our first approach leverages segmentation model, feature extraction and gradient boosted trees. The second approach directly predicts the cellularity from the raw data. Finally, in our third setting, we combine the first and the second approach and used a deep convolutional neural network to predict the cellularity score from segmentation mask.
    }\label{fig:pipeline}
\end{figure*}

Breast cancer is one of the most common cancer types diagnosed in women in the United States and worldwide~\cite{siegel2018stats}. Biopsies and histological assessment allow pathologists to analyze microscopic structures of breast tissues and, in particular, assess the cancer's aggressiveness.

Multiple options are available to manage and monitor the breast cancer treatment based on the information provided from the tumor's response to it. In addition to the treatment effect on the tumor size, the therapy may also alter the tumor's cellularity~\cite{cellularity}. During anticancer therapy, the size of the tumor may remain the same, but the overall cellularity may be drastically reduced~\cite{rajan2004change}. As a result, it makes the residual tumor cellularity an important factor in assessing the response treatment.

Currently, tumor cellularity is manually assessed by pathologists from hematoxylin and eosin (H$\&$E)-stained slides  \cite{rajan2004change}. The costs of such estimation are high, the process is tedious and subjective, and the quality and reliability might be also be affected by high inter-observer variability even among senior pathologists. This potentially may affect prognostic power assessment in clinical trials \cite{symmans2007measurement}. The subjectivity in visual tissue assessment motivates the use of computer-aided methods to improve the diagnosis accuracy, reduce human error and increase inter-observer agreement and reproducibility~\cite{meyer2005breast, elmore2015diagnostic}. Automated analysis of the H\&E slide using computer vision could provide immediate benefits to patient care. Recent success in Deep Learning (DL) \cite{lecun2015deep,schmidhuber2015deep}, and in particular the advances in convolutional neural networks (CNN), have recently shown high potential in this realm~\cite{ching2017opportunities}.

In this work, we evaluate three DL-methods to score the cellularity of the breast tissue from histopathological images. In particular, our first approach employs a weakly-supervised segmentation model with Resnet-34 \cite{he2016deep} encoder and Feature Pyramid Network (FPN)\cite{lin2017feature} and a second-stage regression network that predicts the cellularity score using the predicted segmentation maps. Our second approach is also based on segmentation, however, instead of using the segmentation maps directly, we extract various features from them and use the gradient boosting trees (GBT) \cite{natekin2013gradient} to predict the cellularity score. Finally, we also evaluate using H\&E image patches directly to predict the cellularity score.

\section{Related work}\label{sec:related}

CNNs have recently been successfully applied to many tasks in biomedical image analysis, often outperforming conventional machine learning methods~\cite{ching2017opportunities,tiulpin2018automatic,shvets2018automatic}. As such, they have successfully been utilized for digital pathology image analysis and have demonstrated great potential for improving breast cancer diagnostics~\cite{spanhol2016breast, araujo2017classification, bejnordi2017diagnostic, ROBERTSON2017digital}.

Although there are not many studies focusing directly on automated quantitative cellularity assessment, it has been shown that this task can be solved by first segmenting malignant cells and then computing the tumor's area \cite{peikari2017automatic}. Many efforts have been devoted to developing supervised and unsupervised methods for automated cell and nuclear segmentation and detection~\cite{xing2016robust,komura2018machine}. Supervised segmentation models have superior performance but require hand-labeled nuclear mask annotations~\cite{xing2016robust}. In these approaches, segmented nuclear bodies are used to extract features that are typically inspired by visual markers recognized by pathologists. Commonly used features describe morphology, texture, and spatial relationships among cell nuclei in tissue~\cite{peikari2017automatic,kalinin20183d}. 

The conventional approach most relevant to our work is by Peikari \textit{et al.}~\cite{peikari2017automatic} who proposed an automated cellularity assessment protocol. First, they used smaller patches, or regions of interest (RoI), extracted from whole slide images to segment all present cell nuclei. Then they extracted a number of predefined features from segmented nuclei and used support vector machines to distinguish lymphocytes and normal epithelial nuclei from malignant ones. Cellularity estimation was done using distinguished malignant epithelial figures for every RoI.

Alternatively, segmentation-free methods that directly estimate cellularity from histopathology imaging data and nuclei locations annotated by human observers were also shown promising. In particular, Veta \textit{et al.}~\cite{veta2016cutting} proposed a deep learning-based method that leverages an information from a tumor's cells nuclei locations (centroids) and predicts the areas of individual nuclei and mean nuclear area without the intermediate step of nuclei segmentation. In particular, this approach was based on a 10-layer deep neural network predicting nuclear areas quantized into 20 histogram bins. The results showed that predicted measurements had substantial agreement with manual measurements, which suggests that it is possible to compute the areas directly from imaging data, without the intermediate step of nuclei segmentation. This is in spirit similar to one of our approaches, but we do not directly compare our methods to Veta \textit{et al.} since we use different datasets and performance metrics.

\begin{figure*}[ht]\centering
    \IfFileExists{figures/fpn-net-crop.pdf}{}{\immediate\write18{pdfcrop figures/fpn-net.pdf}}
    \includegraphics[width=\linewidth]{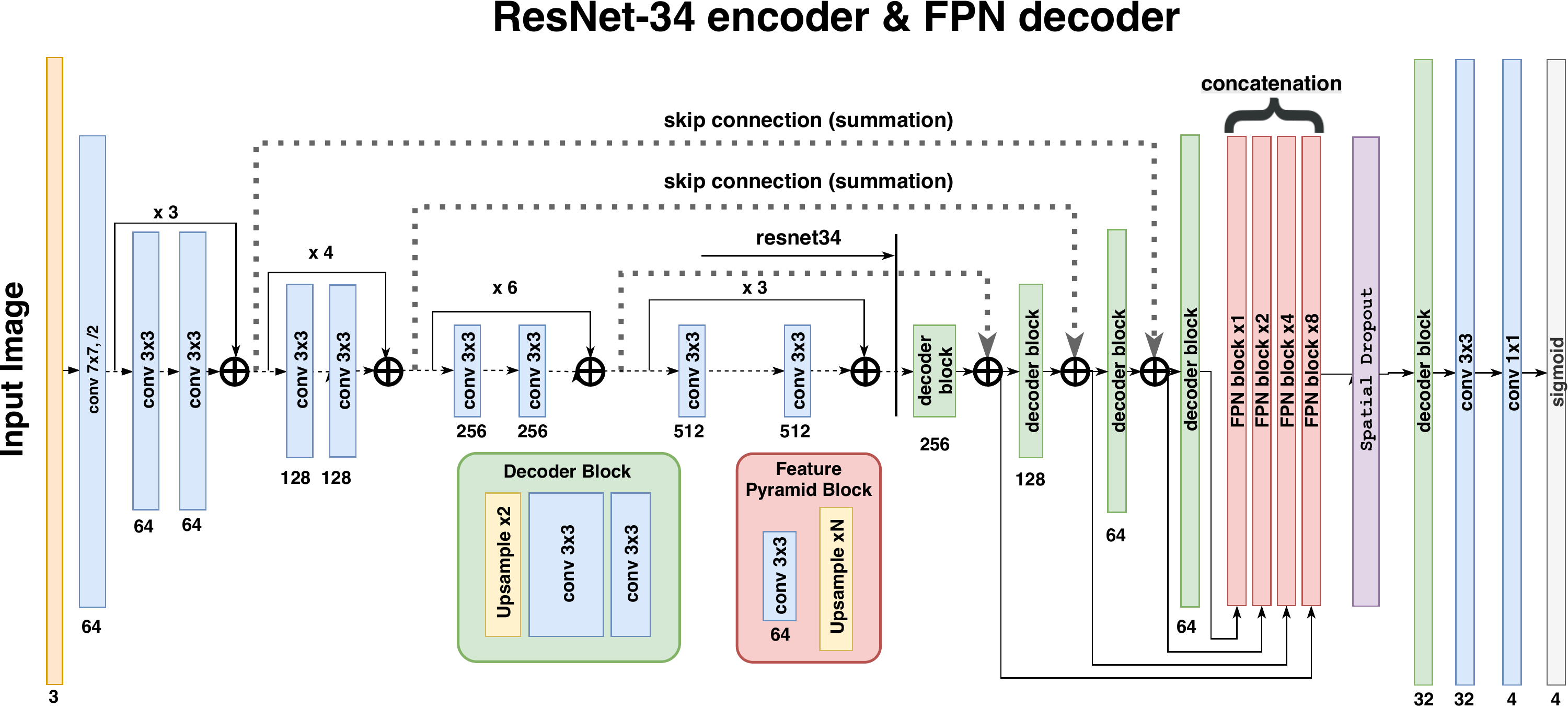}
    \caption{Encoder-decoder segmentation network architecture with Resnet-34 encoder and feature pyramid network decoder. Spatial Dropout 2D is added after multi-layer concatenation.}\label{fig:segmentation_model}
\end{figure*} 

Recent works by Akbar \textit{et al.}~\cite{akbar2019automated,akbar2018determining} have compared the conventional approach based on segmentation and feature extraction and direct applications of deep CNNs to image patches in both regression and classification settings. Overall, they showed that the DL-based approach outperformed hand-crafted features in both accuracy and intra-class correlation (ICC) with expert pathologist annotations. Specifically, their best result was achieved by using a pre-trained Inception \cite{szegedy2016rethinking} model that reached ICC of $0.83$ and $0.81$ with two expert pathologists. In this study we evaluate even wider range of DL-based approaches, including segmentation-based and segmentation-free, in both regression and classification settings. We provide appropriate performance comparisons with previously reported results\footnote{It is worth noting that the official BrestPathQ challenge results have been reported only as a score distribution. Each team know  their  own  results  only, and ours belong to the  right end of the distribution, but, unfortunately,  we are not able to provide a comparison of our approach  with other participants. \url{http://spiechallenges.cloudapp.net/competitions/14\#learn_the_details}}. All the methods developed in this study are fully automatic and do not require any involvement of the human annotators at the test time.

\section{Methods}\label{sec:method}

In this study, we propose and evaluate three different methods. The first two methods are based on the nuclei segmentation and the third method leverages the raw image without preceding segmentation step. Graphical illustration of our approach is presented in Figure~\ref{fig:pipeline}

\subsection{Segmentation}
\paragraph{Network Architecture.} Most modern segmentation architectures inherit the encoder-decoder architecture similar to U-Net~\cite{ronneberger2015unet}, where convolutional layers in the contracting branch (encoder) are followed by an upsampling branch that brings segmentation back to the original image size (decoder). In addition, skip connections are used between contracting and upsampling modules to help the localization information propagate through the complex multilayer structure and eventually improve segmentation accuracy~\cite{ronneberger2015unet}. U-Net and architectures inspired by this idea have produced state of the art results in various segmentation problems, and many improvements for the architecture and its training protocols have recently been proposed. In particular, Iglovikov \emph{et al.}~\cite{iglovikov2018ternausnet} used batch normalization~\cite{ioffe2015batch} and exponential linear unit (ELU) as the primary activation function and an ImageNet pre-trained VGG-11 network~\cite{simonyan2014vgg} as an encoder. Liu \emph{et al.}~\cite{liu2017hourglass} proposed an hourglass-shaped network (HSN) with residual connections, which is also very similar to the U-Net architecture. Rakhlin \emph{et al.}~\cite{Rakhlin_2018_CVPR_Workshops} used the Resnet-34 network~\cite{he2016resnet} as the encoder and the Lov{\'a}sz-Softmax loss function~\cite{Berman_2018_CVPR} along with Stochastic Weight Averaging (SWA)~\cite{izmailov2018averaging} for training.

In our proposed architecture, the segmentation module also inherits the U-Net architecture. The contracting branch (encoder) of our model is based on the Resnet-34~\cite{he2016resnet} network architecture where we have introduced several useful modifications. In particular, we have replaced ReLU activations with ELU that does not saturate gradients and keeps the output close to zero mean and have changed order of batch normalization~\cite{ioffe2015batch} and activation layers. In Section~\ref{sec:eval} we compare encoders initialized with random \textit{He's initialization}~\cite{he2015delving} and pretrained on ImageNet.

\begin{figure*}[ht!]
\centering
\subfloat{\includegraphics[width=0.18\linewidth]{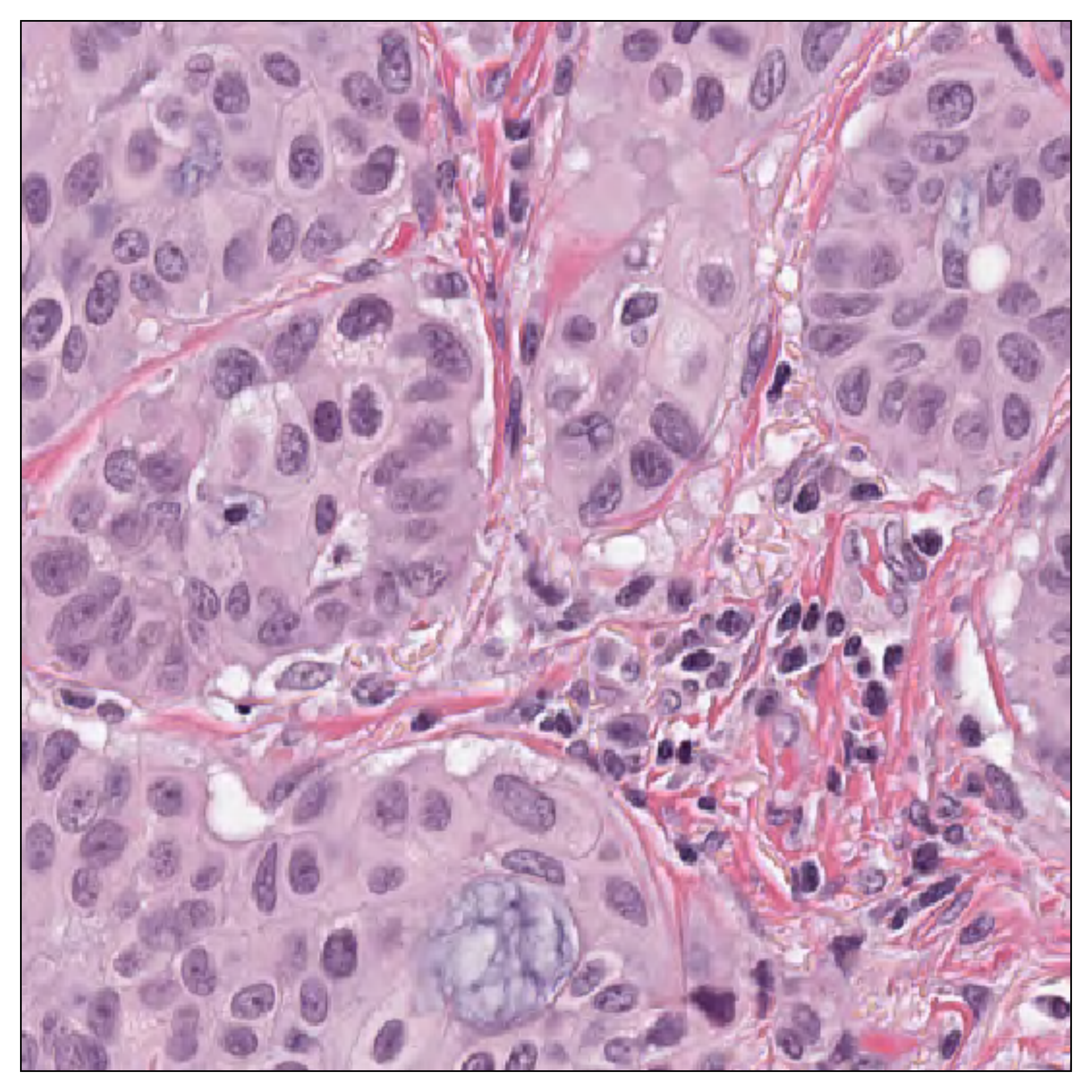}}
\hfill
\subfloat{\includegraphics[width=0.18\linewidth]{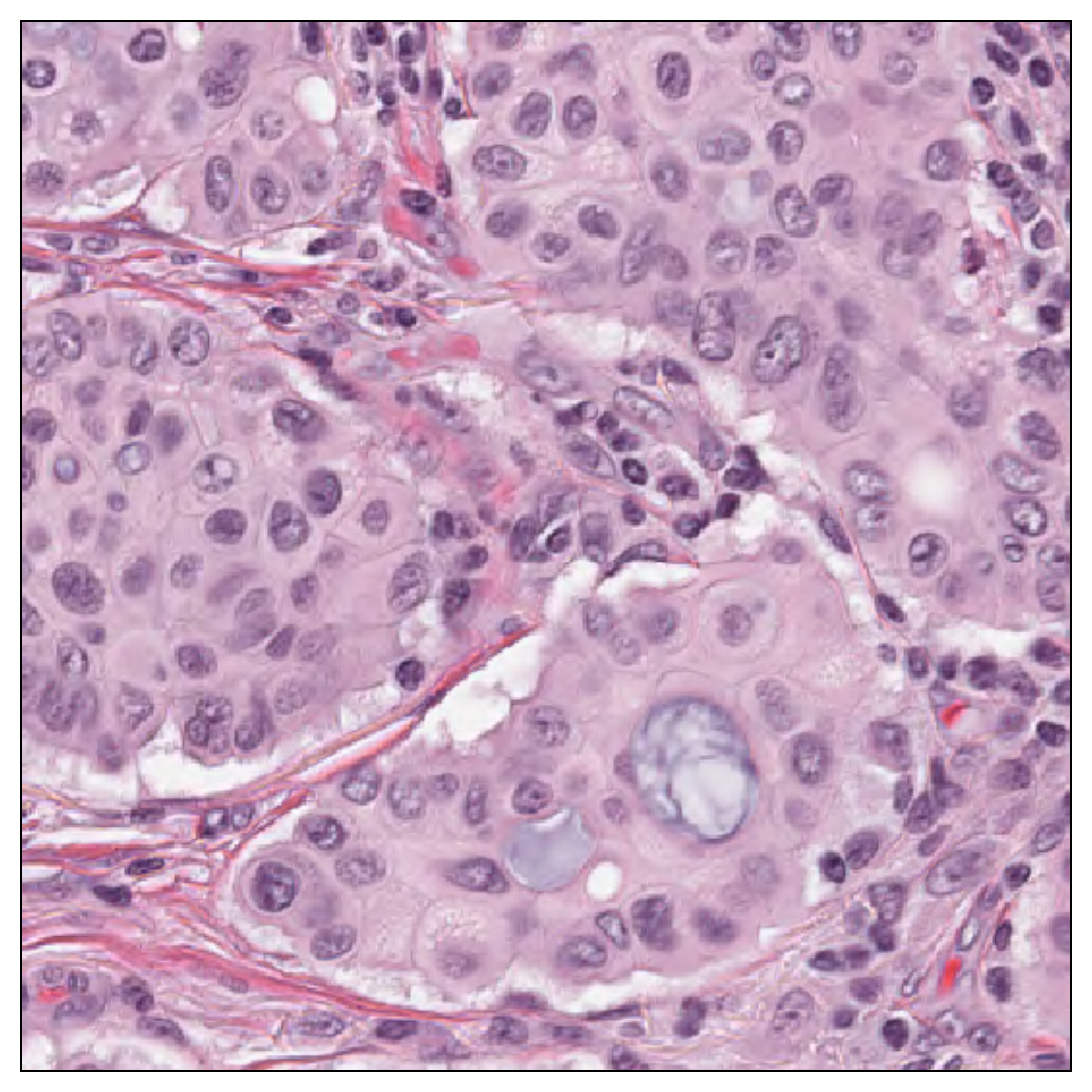}}
\hfill
\subfloat{\includegraphics[width=0.18\linewidth]{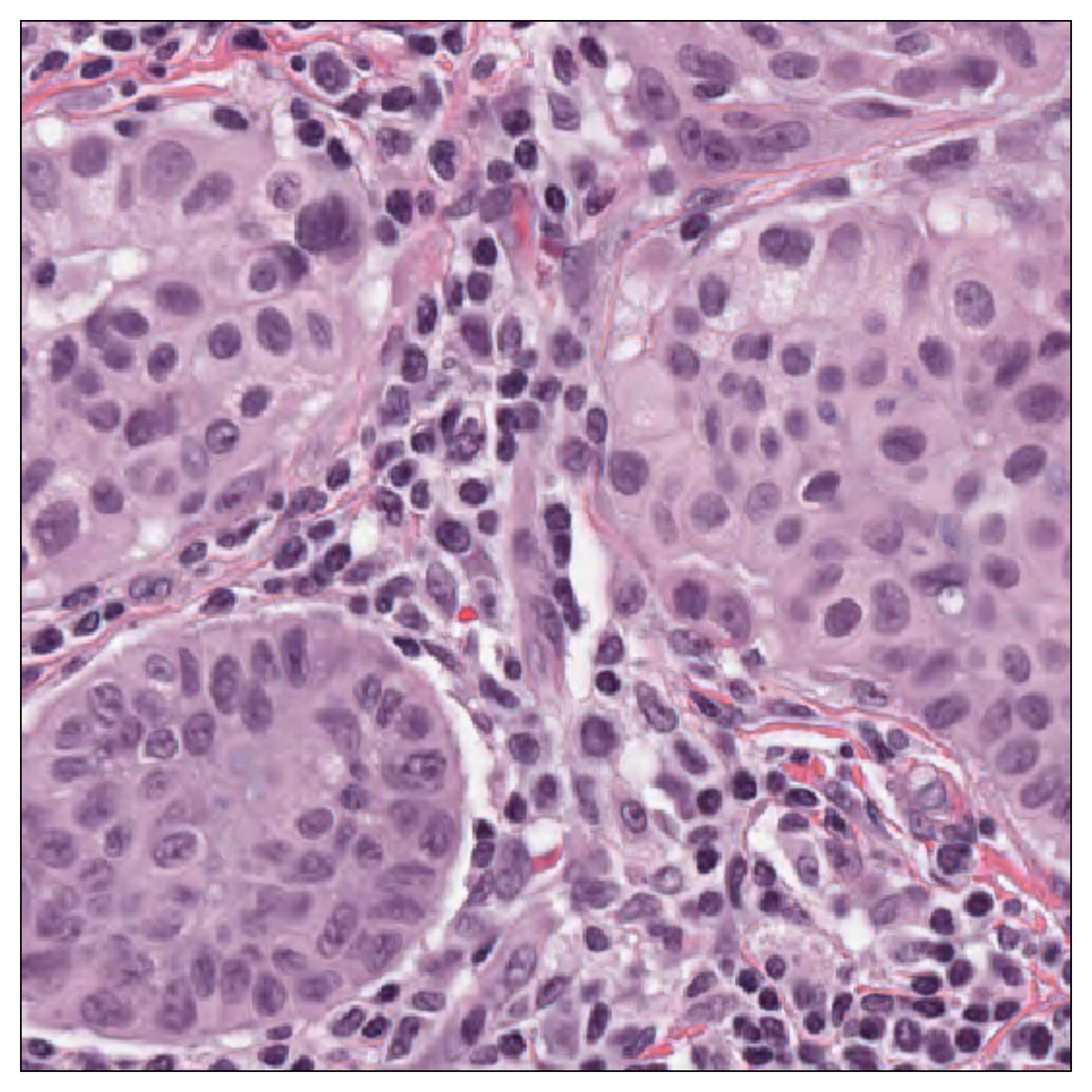}}
\hfill
\subfloat{\includegraphics[width=0.18\linewidth]{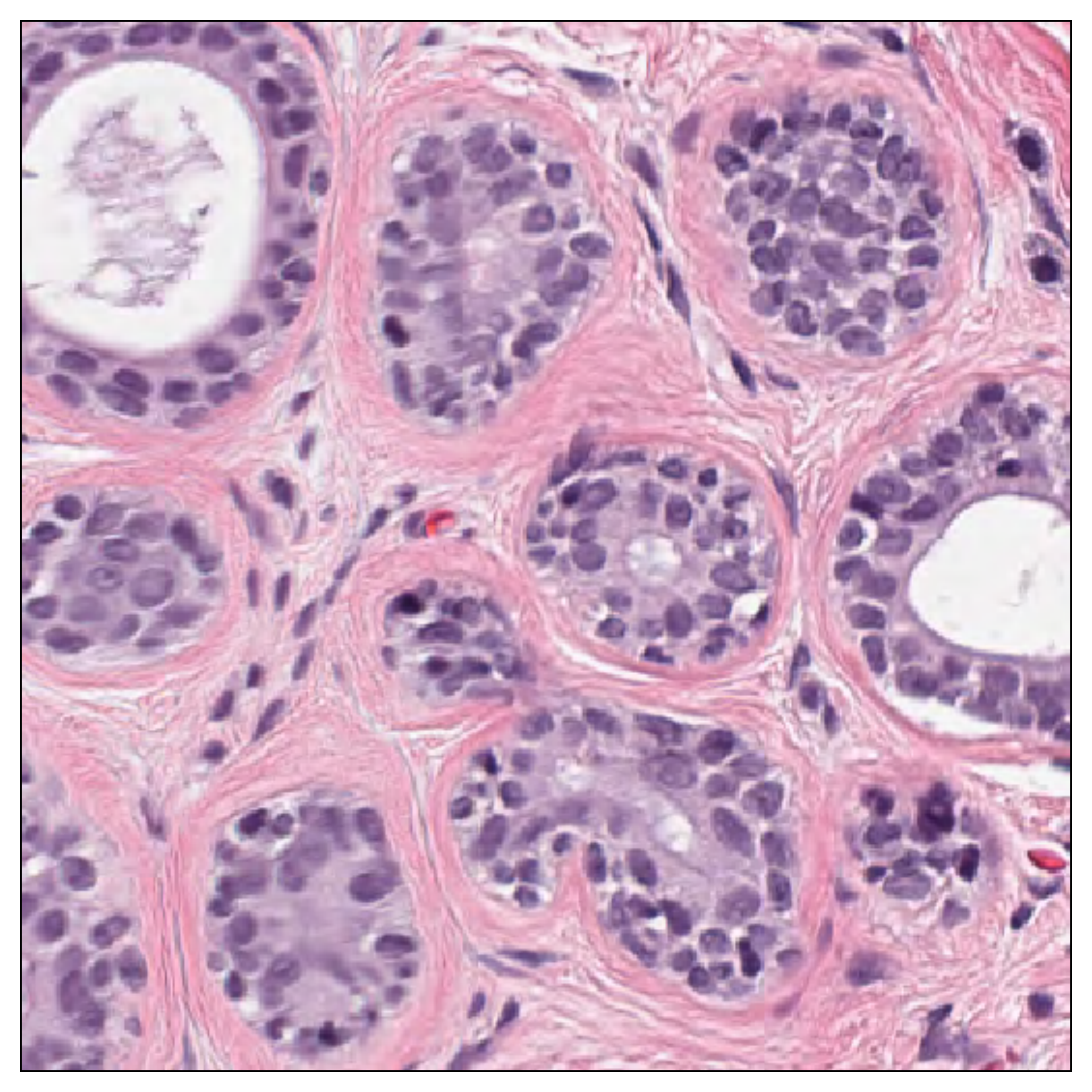}}
\hfill
\subfloat{\includegraphics[width=0.18\linewidth]{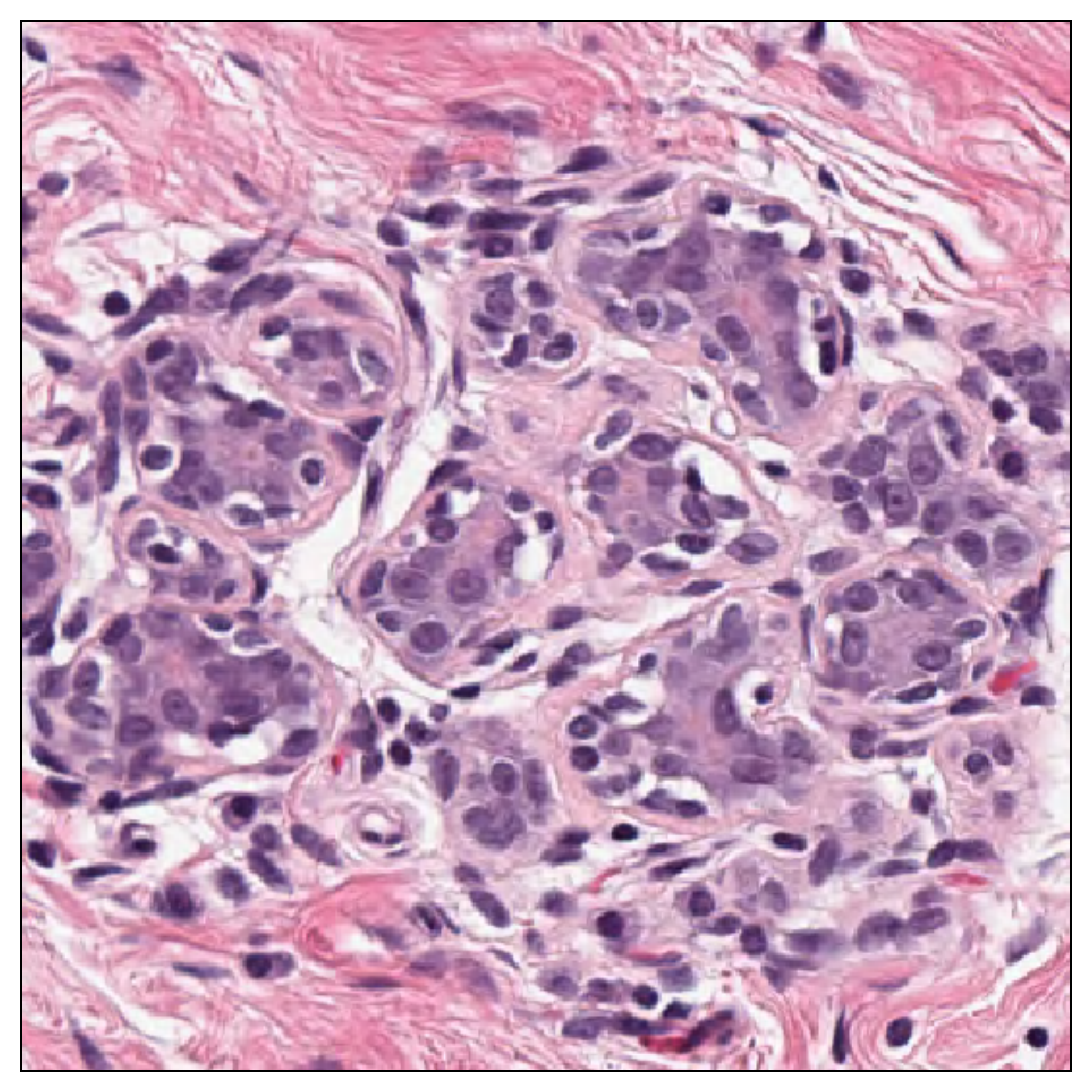}}
\setcounter{subfigure}{0}
\subfloat[]{\includegraphics[width=0.18\linewidth]{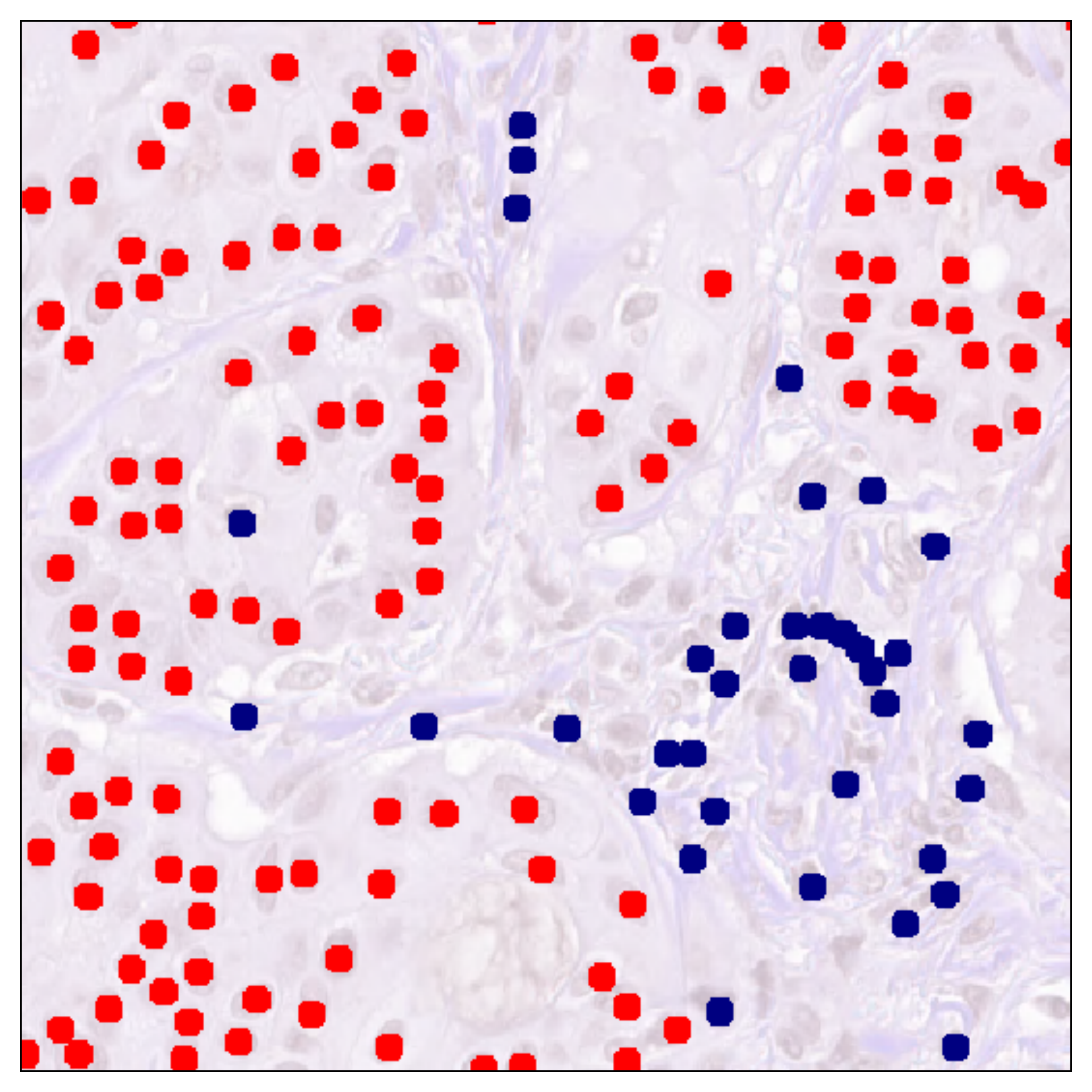}}
\hfill
\subfloat[]{\includegraphics[width=0.18\linewidth]{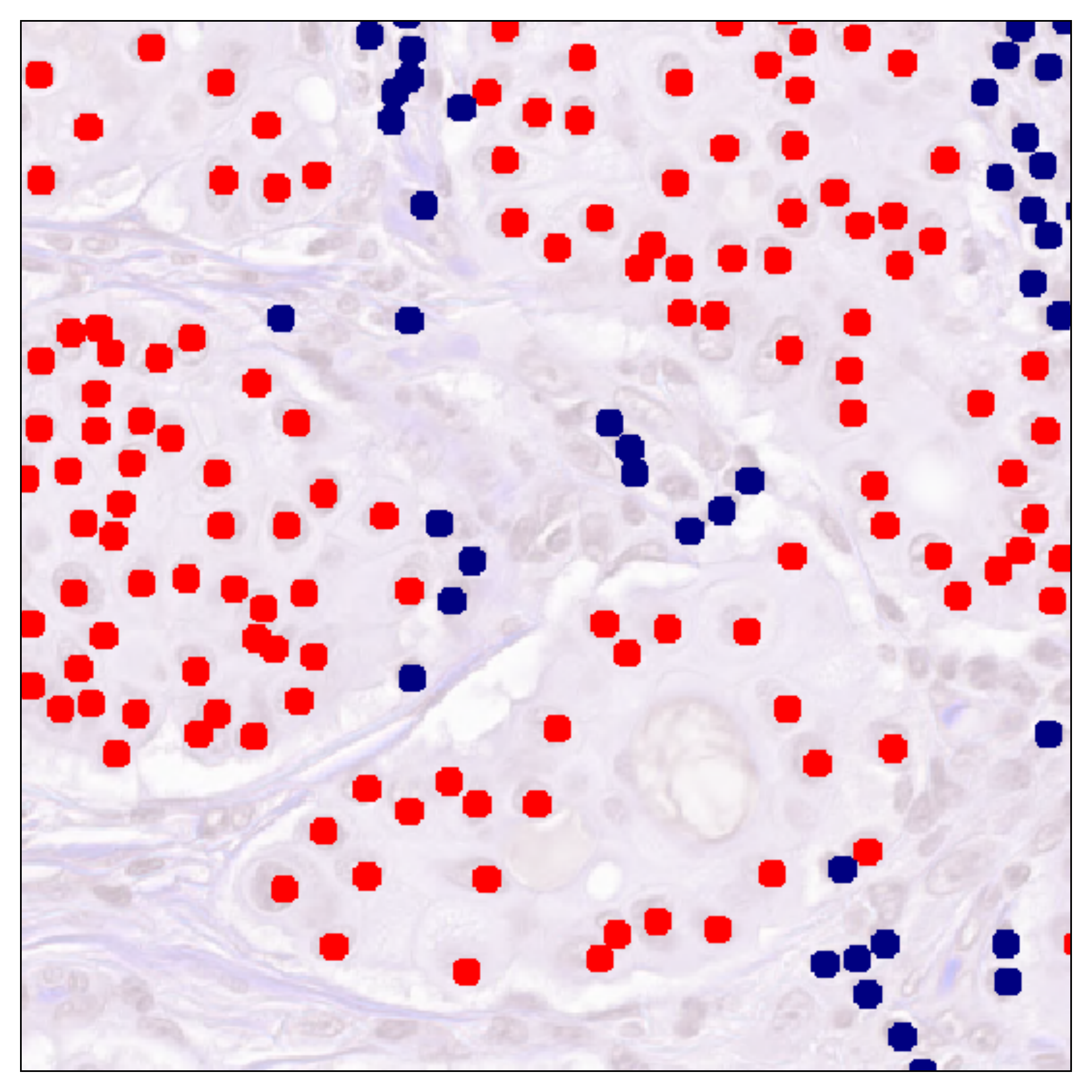}}
\hfill
\subfloat[]{\includegraphics[width=0.18\linewidth]{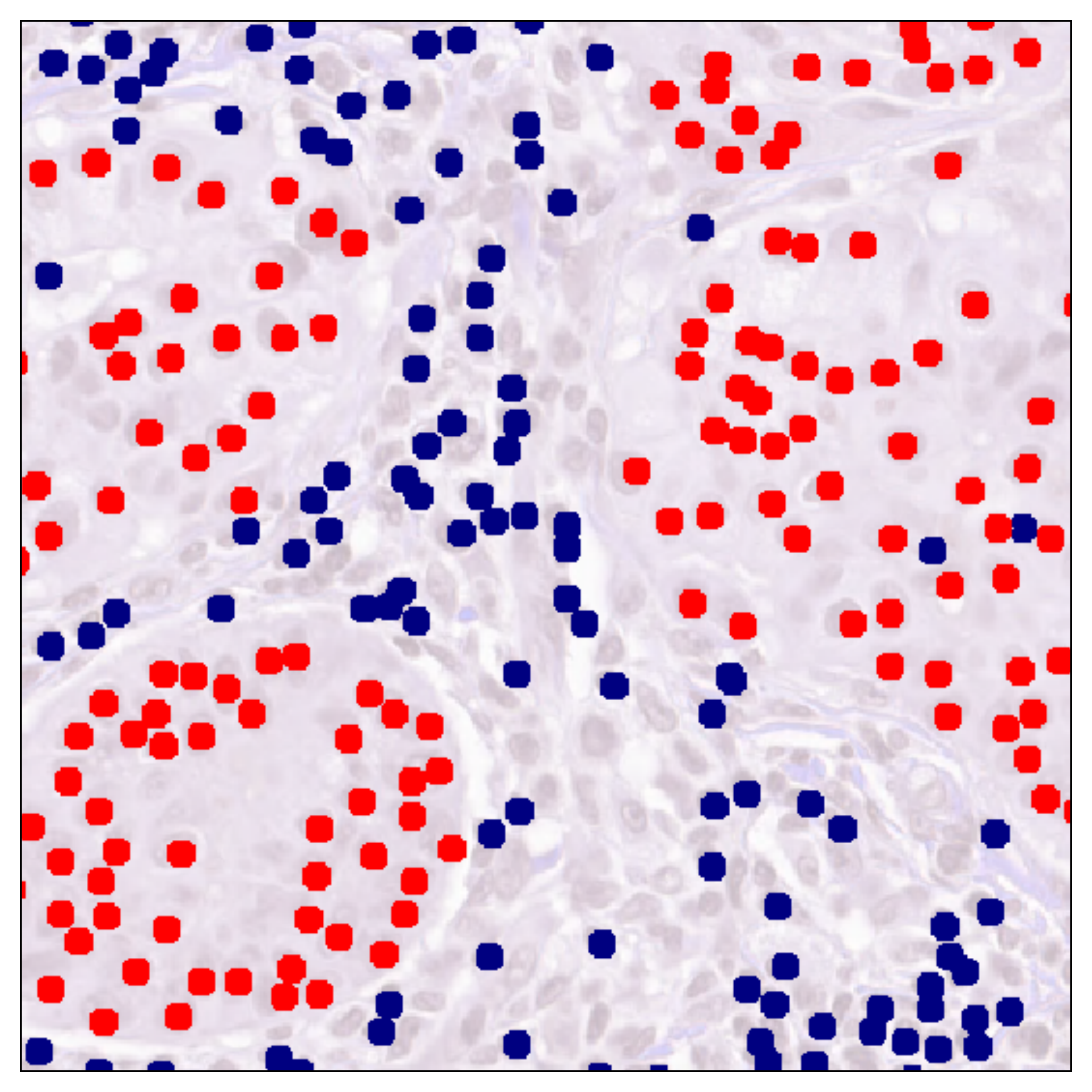}}
\hfill
\subfloat[]{\includegraphics[width=0.18\linewidth]{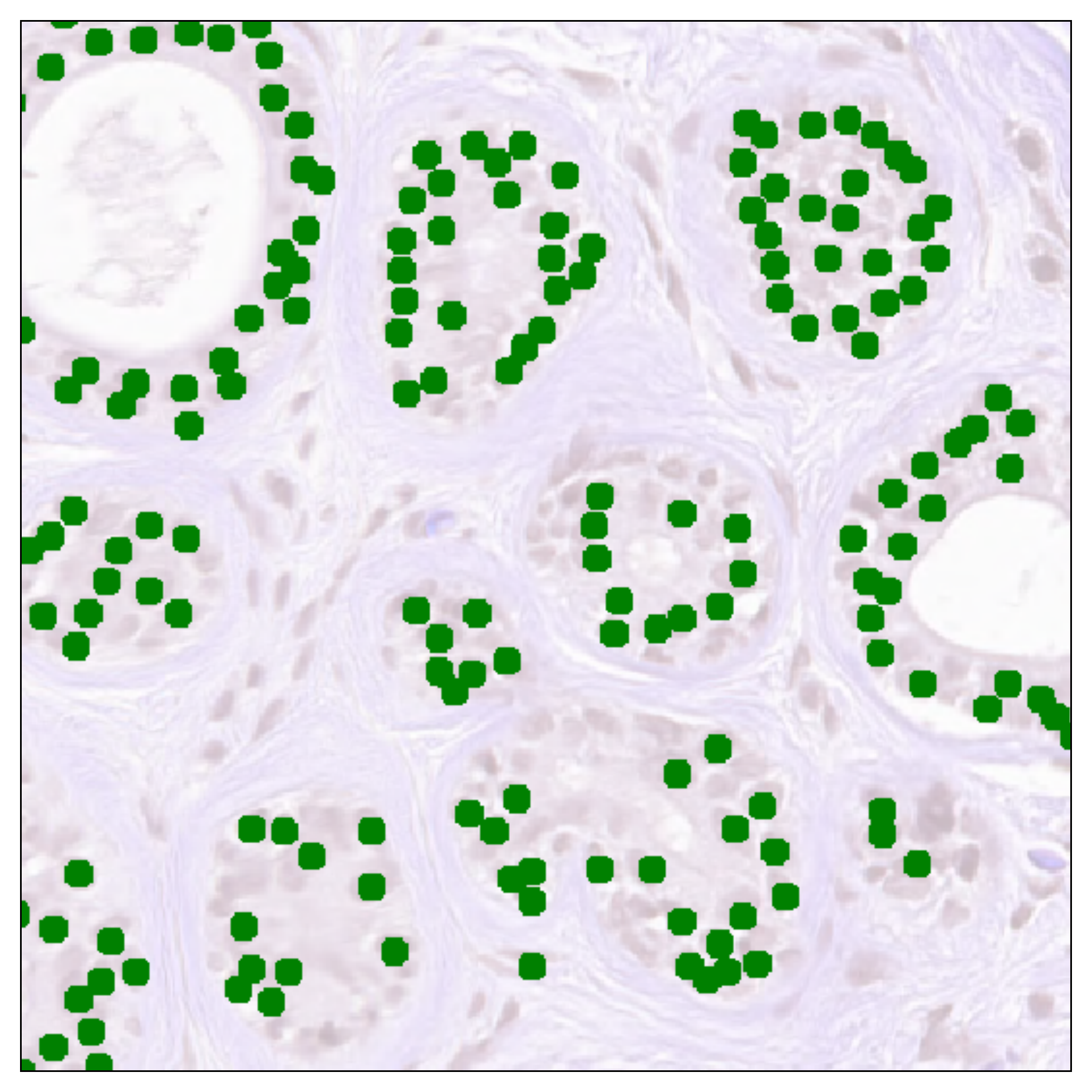}}
\hfill
\subfloat[]{\includegraphics[width=0.18\linewidth]{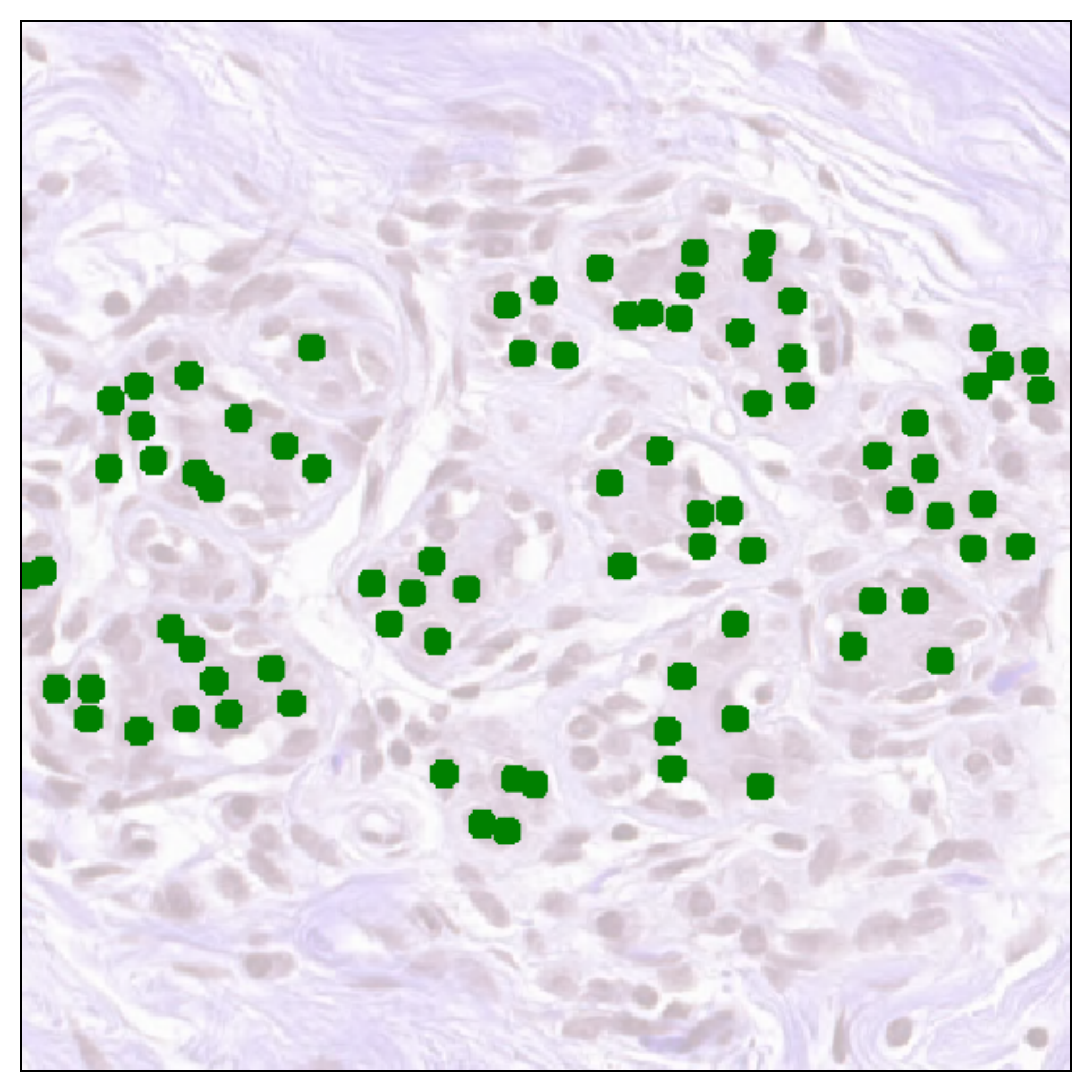}}

\caption{Light micrograph of a histologic specimen of breast tissue stained with hematoxylin and eosin (top). The bottom row shows nuclei segmentation masks synthesized from weak labels: \emph{Malignant}~--- red, \emph{Normal}~--- green, \emph{Lymphocyte}~--- blue.}\label{fig:nuclei}
\end{figure*} 

To address the limited size of the BreastPathQ Cancer Cellularity Challenge dataset, we utilized two regularization techniques: (1) data augmentation and (2) spatial 2D dropout incorporated into the upsampling branch~\cite{tompson2015efficient}.

The upsampling branch is implemented as a Feature Pyramid Network (FPN)~\cite{DBLP:journals/corr/LinDGHHB16}, reconstructing high-level semantic feature maps at four scales simultaneously. We implement a feature pyramid block as a convolutional layer with 64 activation maps followed by upsampling to the original resolution with upsampling rate of 8, 4, 2, or 1 depending on the feature map depth (see Fig.~\ref{fig:segmentation_model}). In Section~\ref{sec:eval}, we compare the performance of standard and FPN decoders.

We concatenate upsampled maps into a single layer of $64\times4=256$ maps and add after it a spatial 2D dropout layer, which acts as a regularizer and prevents coadaptation of the network weights, but unlike conventional dropout it drops out not individual neurons but rather entire activation maps. Throughout the work, we use dropout rate $0.5$, randomly dropping $128$ out of $256$ activation maps.

Finally, the output of the model is a 4-channel sigmoid layer that assigns every pixel with four values from $0$ to $1$ that represent the probabilities of belonging to the \textit{Normal}, \emph{Lymphocyte}, \emph{Malignant}, and \textit{Background} classes.

\paragraph{Loss functions.} Binary cross entropy (BCE), while convenient for training, does not directly translate into Jaccard index, the metric commonly used to evaluate segmentation accuracy. Hence, as the loss function we use 
\begin{equation}
L^c(w)=(1-\alpha) \mathrm{BCE}^c(w) - \alpha J^c(w),
\label{loss}
\end{equation}
a weighted sum of $\mathrm{BCE}$ and the soft Jaccard loss for class $c$ \cite{ iglovikov2018ternausnet, Rakhlin_2018_CVPR_Workshops, iglovikov2017pediatric}. In this work, we set $\alpha=0.15$, a value found via cross-validation. The soft Jaccard loss is defined as
\begin{equation}
J^c(w)=\frac{1}{N}\sum\limits_{i=1}^N\left(\frac{y^c_i\hat{y}^c_i}{y^c_i+\hat{y}^c_i-y^c_i\hat{y}^c_i}\right),
\label{jaccard_loss}
\end{equation}
where $w$ are network parameters, $y^c_i$ is the binary label for pixel $i$ and class $c$, $\hat{y}^c_i$ is the predicted probability of $c$ for pixel $i$, and $N$ is the total number of pixels.
The total loss function is a weighted sum of class losses:
\begin{equation}
L(w)=\frac{1}{V}\sum\limits_{c=1}^4L^c(w)v^c, V=\sum\limits_{c=1}^4v^c,
\label{total_loss}
\end{equation}
where $v^c$ is a loss weight for class $c$. In this work we weigh \emph{Normal}, \emph{Lymphocyte}, and \emph{Background} as $1$ and \textit{Malignant}, the class of primary importance in our problem, as $4$.

\subsection{Cellularity estimation from segmented cells} In this subsection, we describe the method for cellularity assessment that leverages the output of the trained segmentation network Fig.~\ref{fig:segmentation_model}. We feed the segmented output into a Resnet-34 CNN model. The model automatically learns deep features from the 4-channel segmentation input and regresses it onto continuous cellularity score using continuous regression loss ($L_2)$. In this approach, the segmentation model acts as a filter the aim of which is to extract only the information about the cell morphology. We hypothesized that this structured approach makes our method similar to methods employed by expert pathologists, makes it transparent and less sensitive to data acquisition settings.

\subsection{Feature extraction-based cellularity estimation}

\begin{figure}[!t]\centering
\subfloat[$0.04$]{\includegraphics[width=0.25\linewidth]{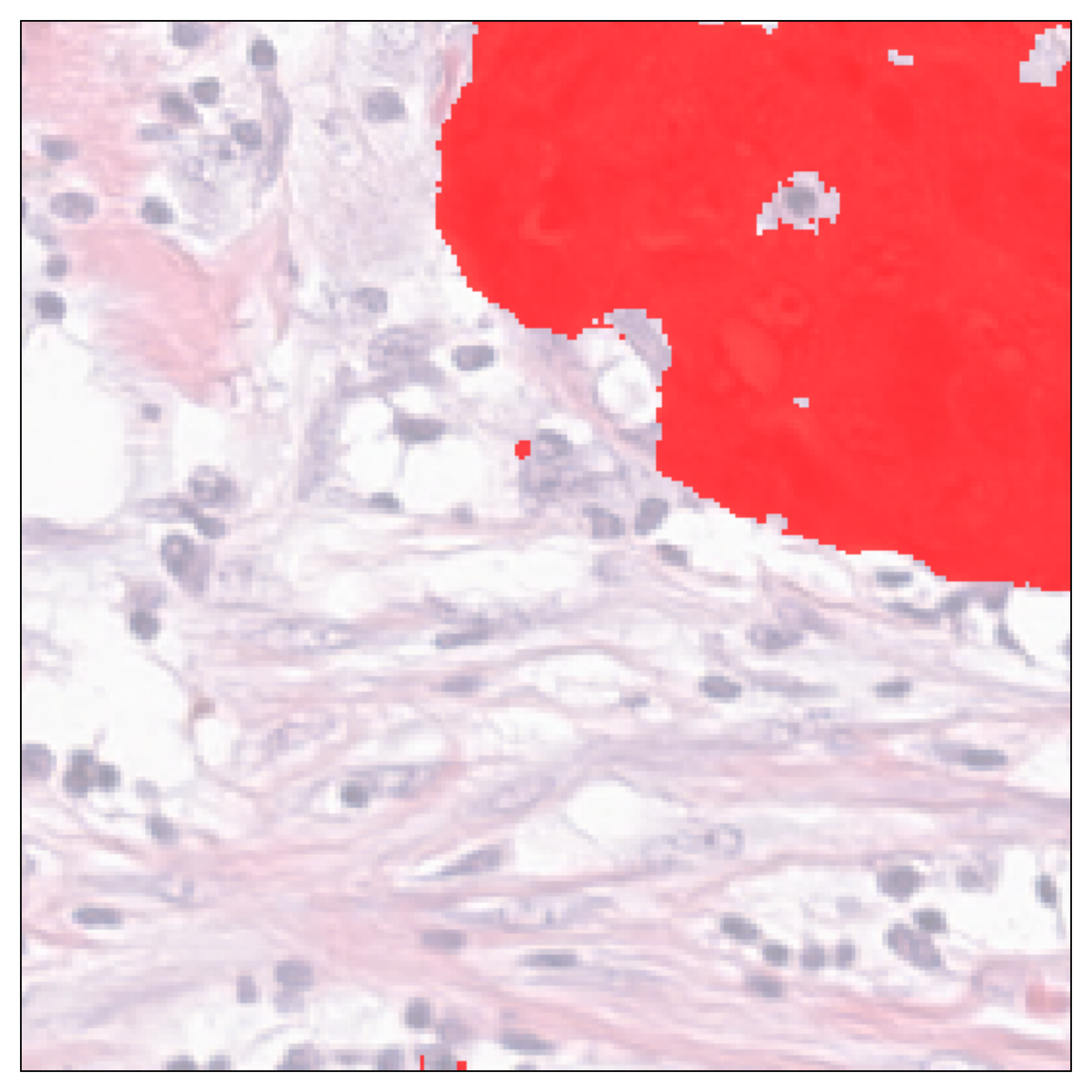}}
\hfill
\subfloat[$0.16$]{\includegraphics[width=0.25\linewidth]{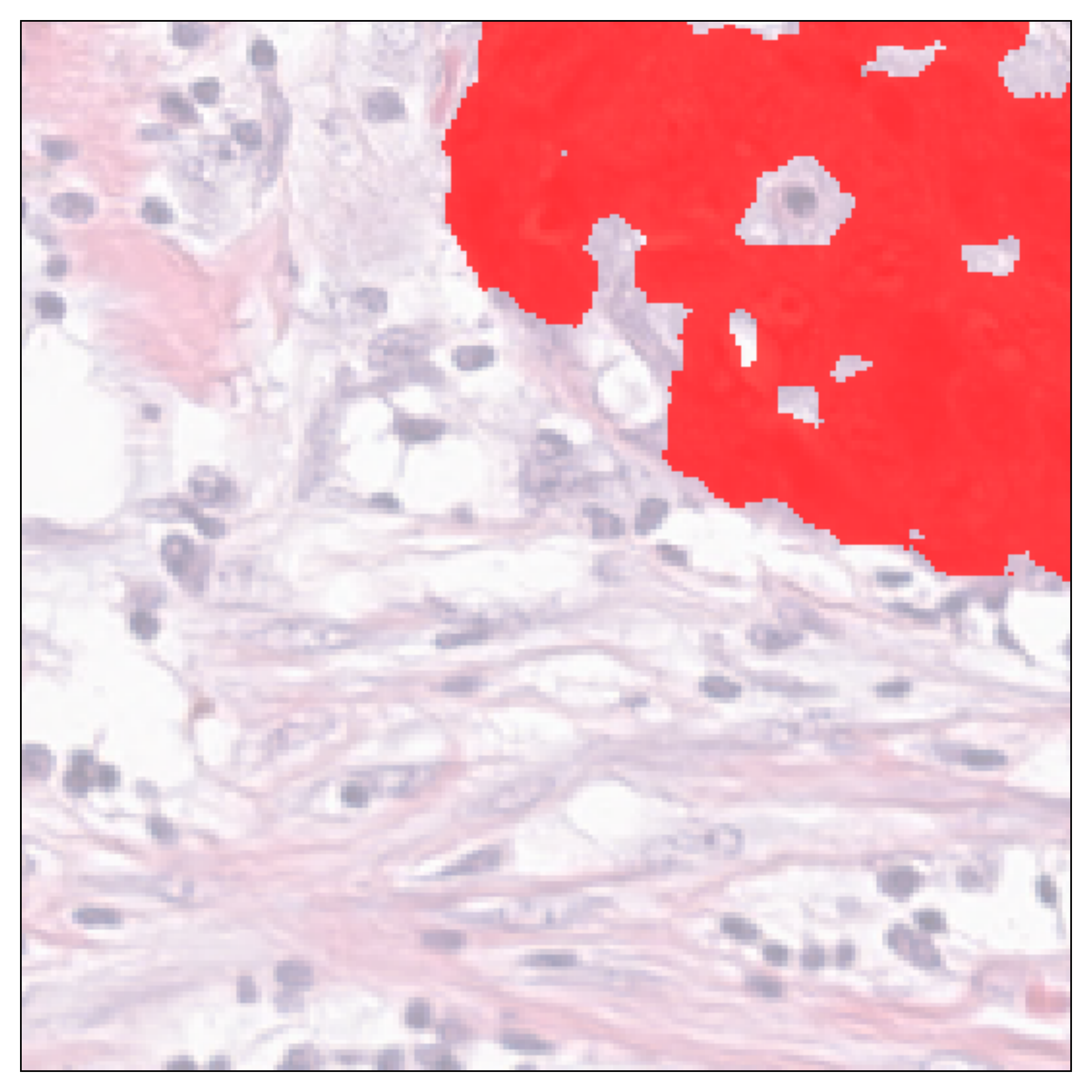}}
\hfill
\subfloat[$0.32$]{\includegraphics[width=0.25\linewidth]{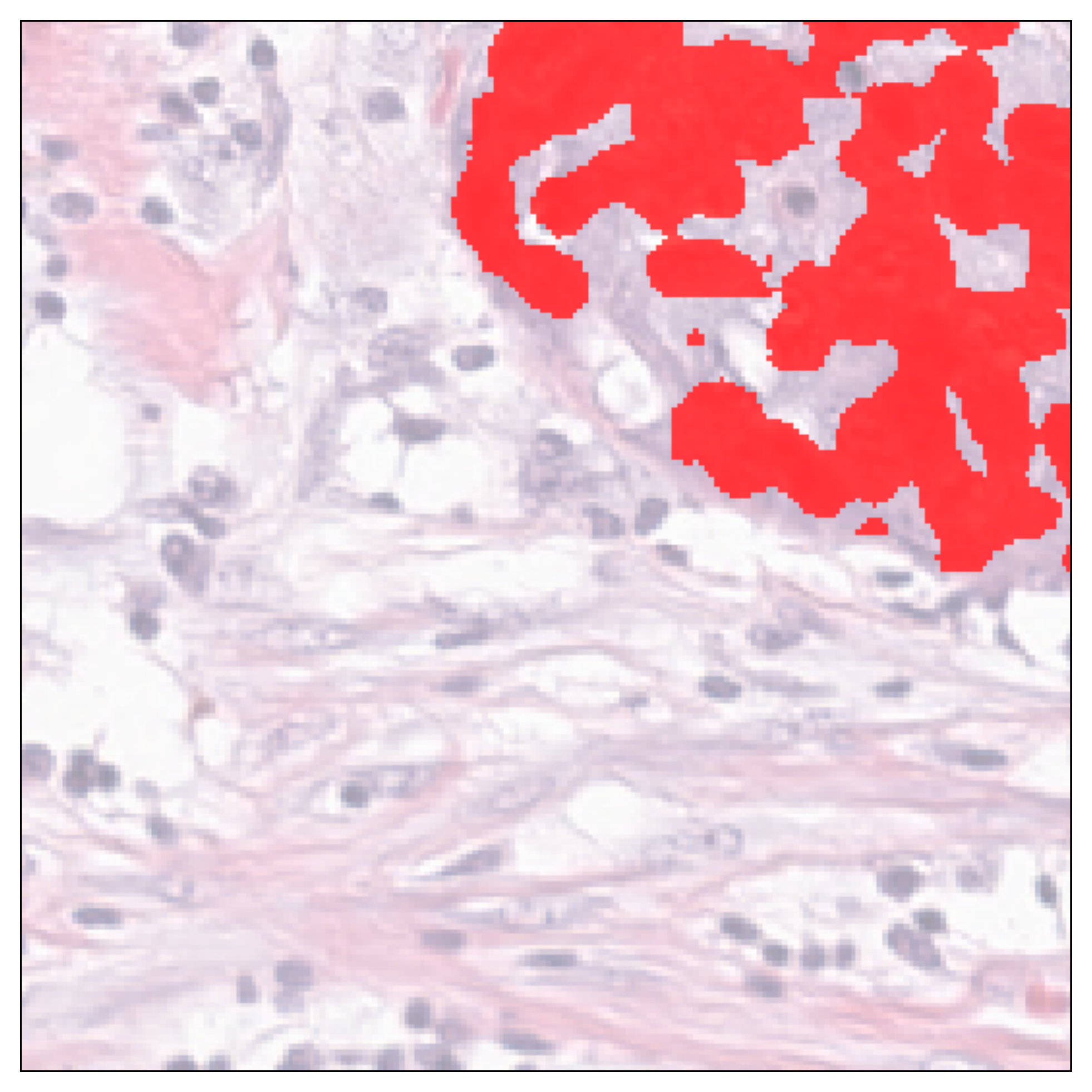}}
\hfill
\subfloat[$0.50$]{\includegraphics[width=0.25\linewidth]{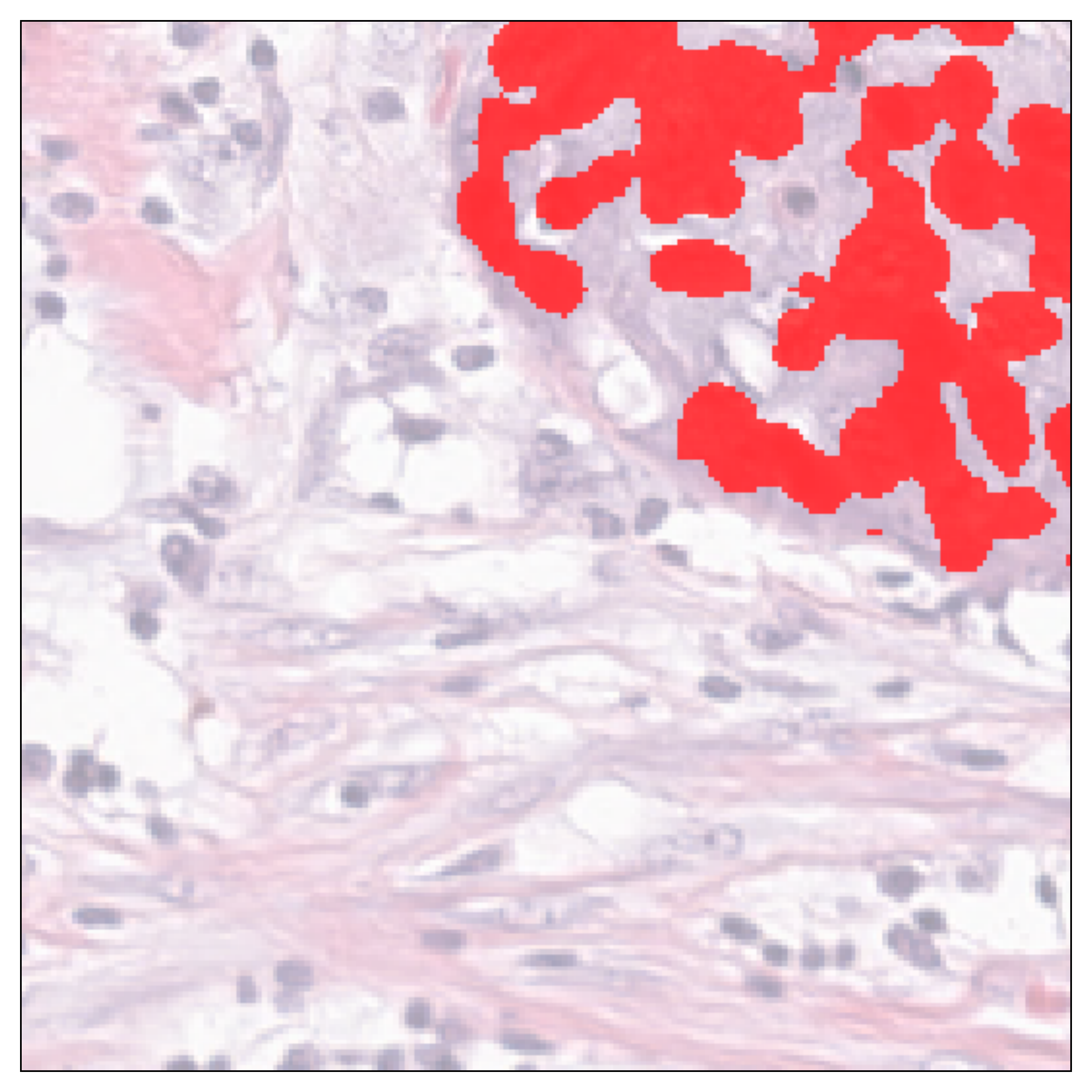}}
\caption{Segmentation results at thresholds (a) - (d) of \emph{Malignant} channel superimposed with the original image. Masks generated after thresholding were used for feature extraction. }\label{fig:1st-type-features}\vspace{.3cm}
\subfloat[$0.02$]{\includegraphics[width=0.25\linewidth]{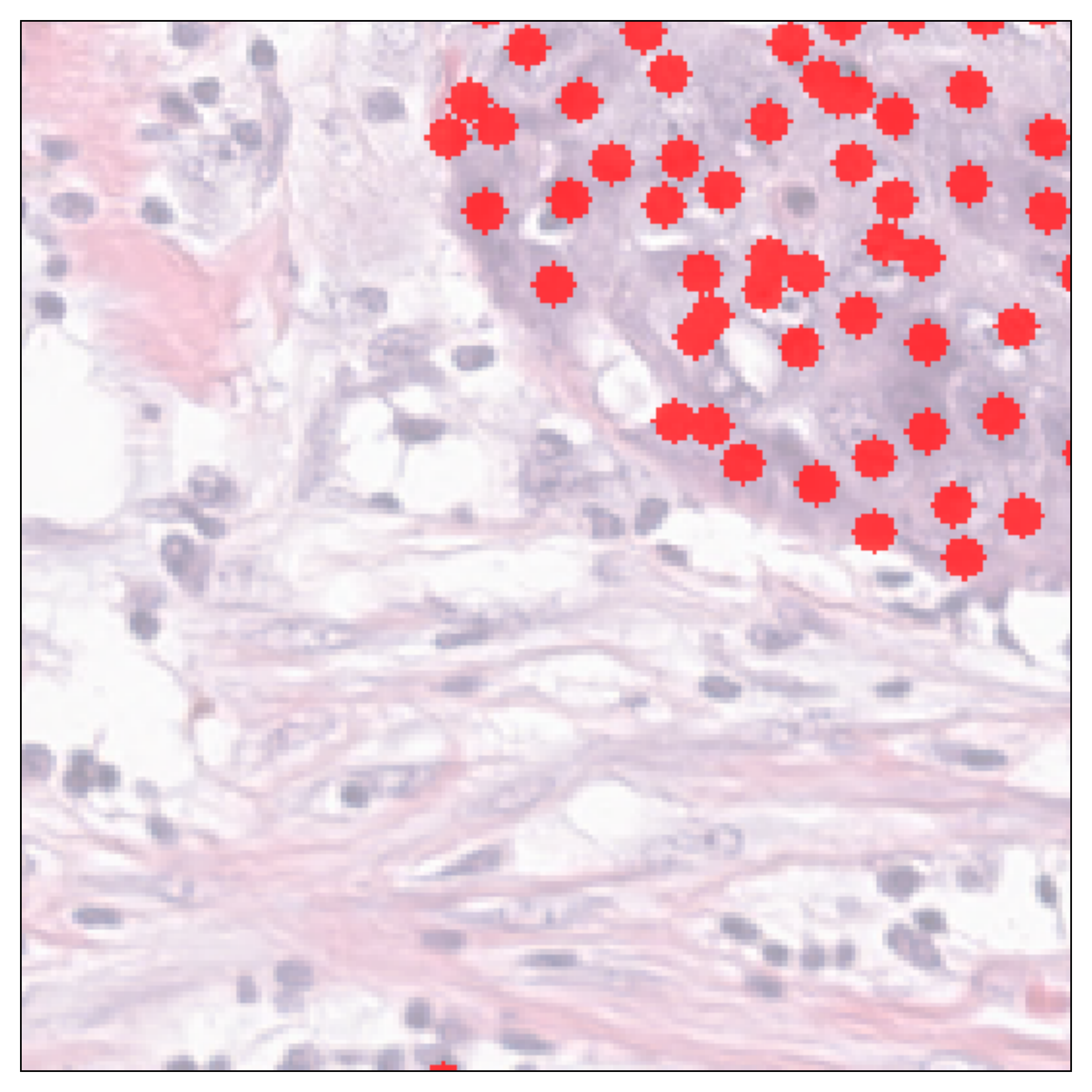}}
\hfill
\subfloat[$0.08$]{\includegraphics[width=0.25\linewidth]{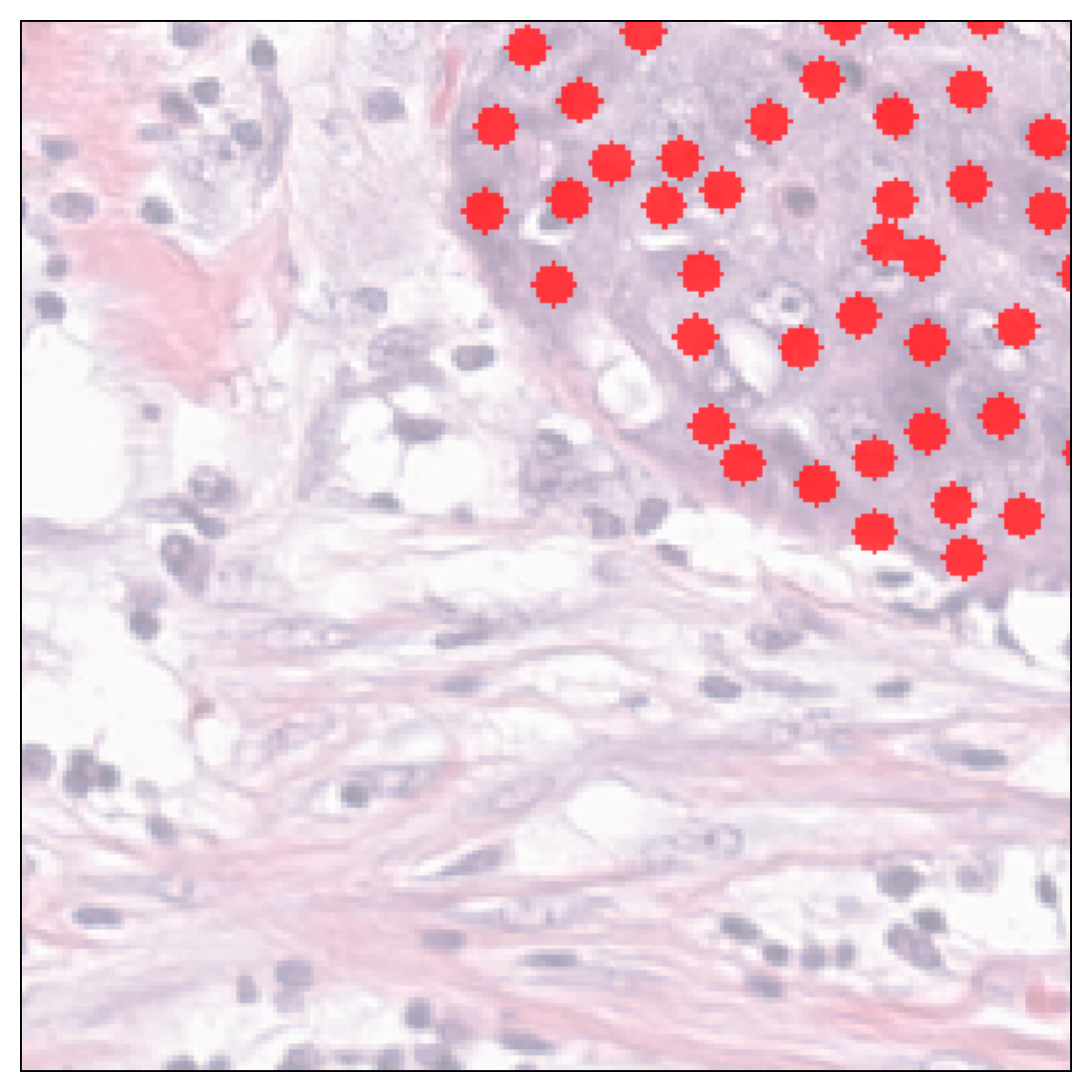}}
\hfill
\subfloat[$0.16$]{\includegraphics[width=0.25\linewidth]{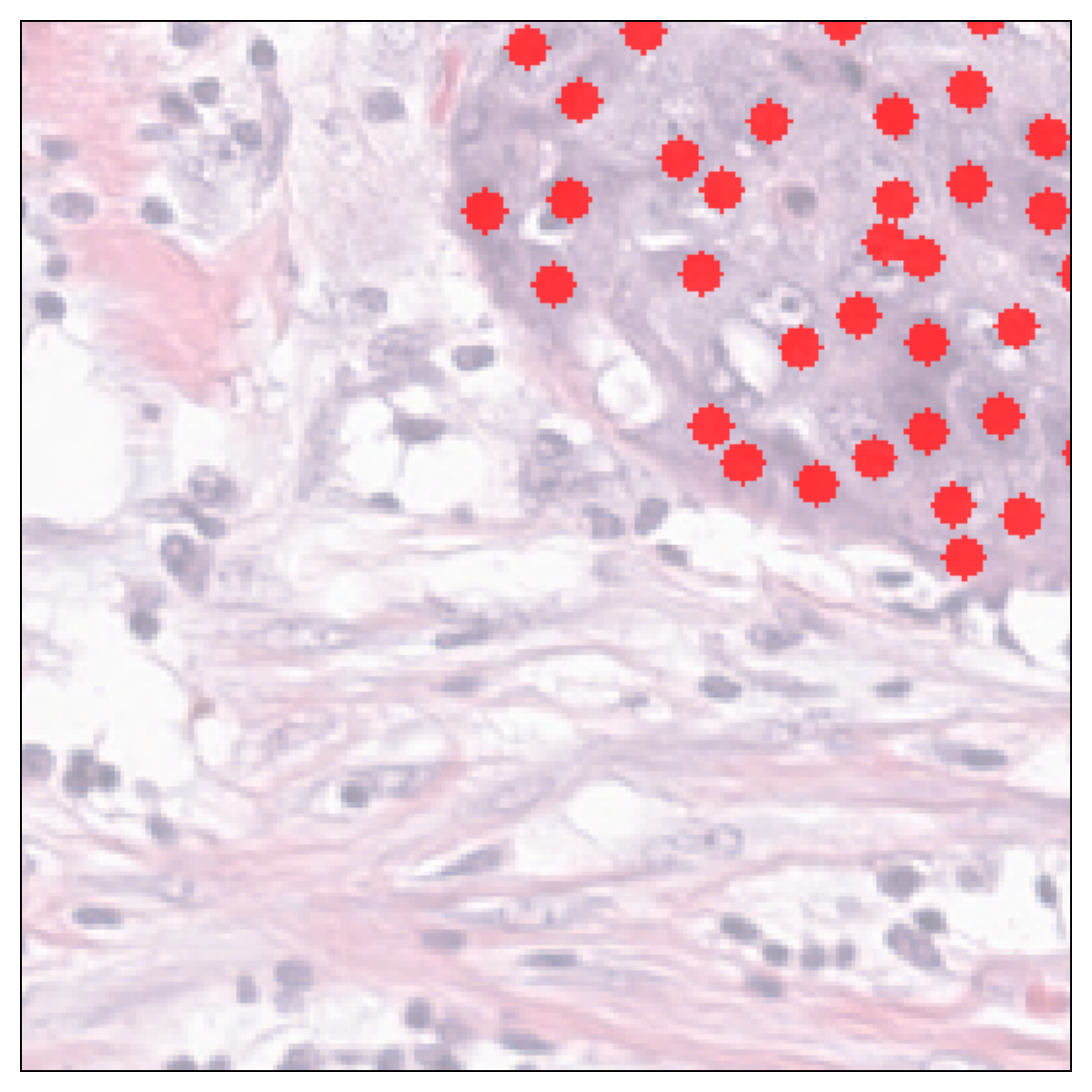}}
\hfill
\subfloat[$0.24$]{\includegraphics[width=0.25\linewidth]{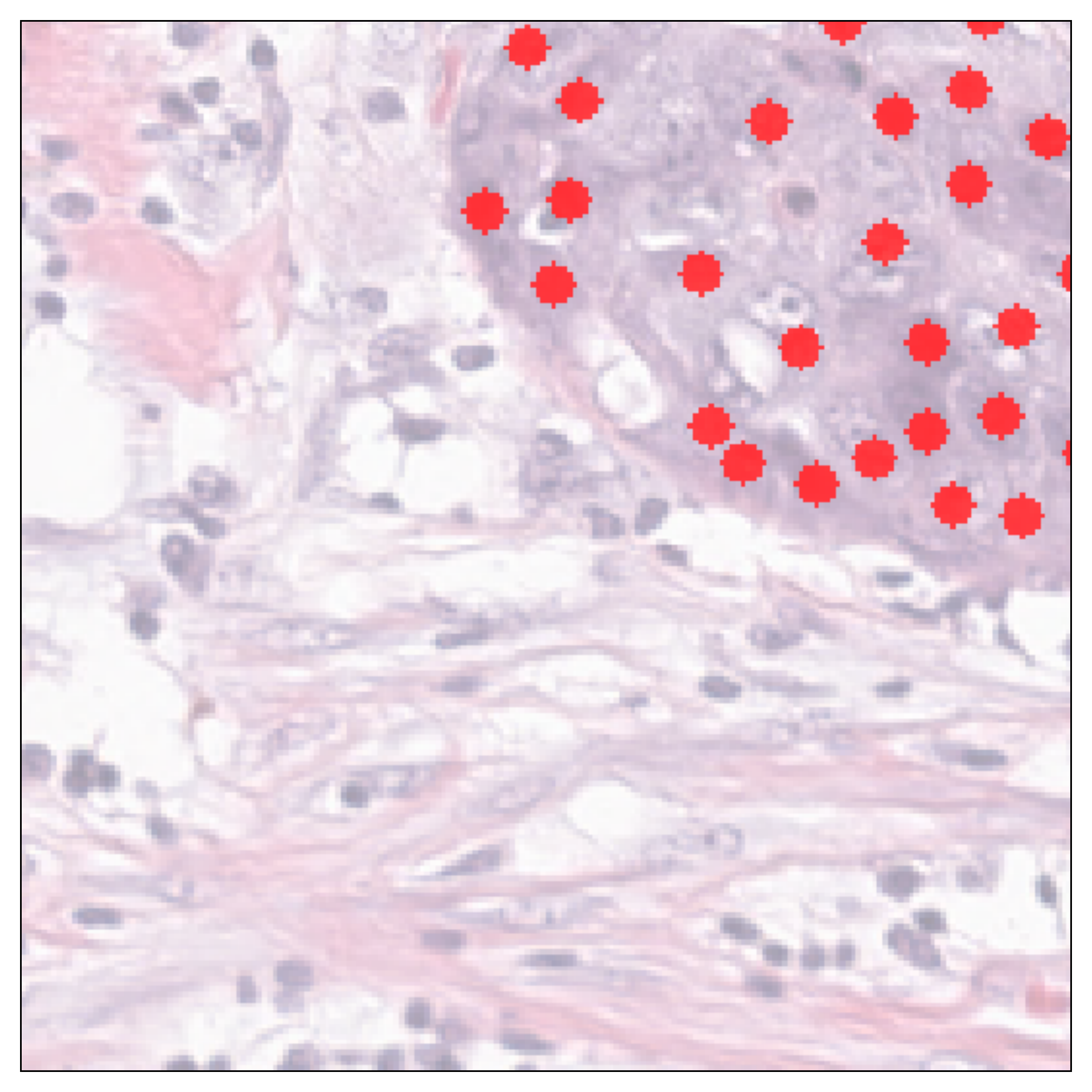}}

\caption{Nuclei blobs detected from the \emph{Malignant} segmentation maps using the Laplacian of Gaussian method at thresholds (a)-(d). The blobs were used for feature extraction.}\label{fig:2nd-type-features}\vspace{-.7cm}

\end{figure}

The second type of model is Gradient Boosted Trees (GBT)~\cite{NIPS2017_6907} in regression mode ($L_2$ loss). The general idea and handcrafted features are borrowed from the second place solution for 2017 Kaggle contest for Sea Lion Population Count in aerial imaginary~\cite{lopuhin}. The authors would like to thank Konstantin Lopuhin for valuable discussion they had while incorporating his method. In this study, GBT operates on a vector of hand-crafted features extracted from nuclei segmentation maps, including:
\begin{itemize}
\item activations and their areas aggregated over segmentation maps with different thresholds; for every segmentation map in \emph{Normal}, \emph{Lymphocyte}, \emph{Malignant} and for 7 thresholds $0.02$, $0.04$, $0.08$, $0.16$, $0.24$, $0.32$, $0.5$, we obtain 2 values: total area above threshold and total activation above threshold (see Fig.~\ref{fig:1st-type-features} for an illustration);
\item using the Laplacian of Gaussian (LoG) method as implemented in the OpenCV library~\cite{bradski2000opencv}, we find blobs in segmentation maps at 6 thresholds: $0.02$, $0.04$, $0.08$, $0.16$, $0.24$, $0.5$; for each threshold we find the number of blobs and total activation in blob centers (Fig.~\ref{fig:2nd-type-features});
\item total activation for every channel, computed as a sum of the activations at every pixel after sigmoid.
\end{itemize}

In total, we obtain $3 \times (7 \times 2 + 6 \times 2 + 1) = 81$ features to train the GBT model.

\subsection{Cellularity estimation from the raw images}

The third type of model is a deep convolutional network implemented in regression or classification settings ($L_2$ or categorical cross-entropy loss functions respectively). These models do not use intermediate segmentation and predict the cellularity score immediately from the microscopic image. For classification, we categorize cellularity into $101$ class using regular bins with thresholds $0.00,0.01,\ldots,1.00$. The idea of direct regression of an image into continuous value using CNN is not new. In particular, it was implemented in~ \cite{iglovikov2017pediatric} where the authors use CNN to predict bone age from radiograph.

\subsection{Evaluation metrics}

We assessed the results using several metrics. The main evaluation metric is the mean squared error (MSE) between the cellularity score obtained in our experiments and ground truth provided by an expert pathologist.

In order to make our results comparable with previous work, we also report Cohen's kappa coefficient agreement and the intra-class correlation coefficient (ICC) between expert and automated methods, similar to~\cite{peikari2017automatic}. In all experiments, we find our results superior to our predecessors; however, the cellularity score itself in~\cite{peikari2017automatic} is evaluated based on binning it into four categories of $0\mbox{--}25$\%, $26\mbox{--}50$\%, $51\mbox{--}75$\%, and $76\mbox{--}100$\%. Such $4$-class categorization is relatively coarse and, in our opinion, does not represent a suitable evaluation metric for continuous cellularity estimation that is our goal in this work.

\section{Experiments and results}\label{sec:eval}

\subsection{Data}

The data used in this study had been acquired from the Sunnybrook Health Sciences Centre with funding from the Canadian Cancer Society and was made available for the BreastPathQ challenge sponsored by the SPIE, NCI/NIH, AAPM, and the Sunnybrook Research Institute~\cite{peikari2017automatic}.

\pgfplotsset{every axis/.append style={ %
    font=\scriptsize\sffamily,
    y label style={at={(axis description cs:0.1,.5)},anchor=south},
    legend style={
        at={(1.0,0.0)},
        anchor=south east,
        legend columns=1,
        draw=none,
    },
    legend cell align=left,
} }

\pgfplotsset{every axis plot/.append style={
	line width=1.pt,
	mark size=1.25pt,
	mark repeat=24,
} }

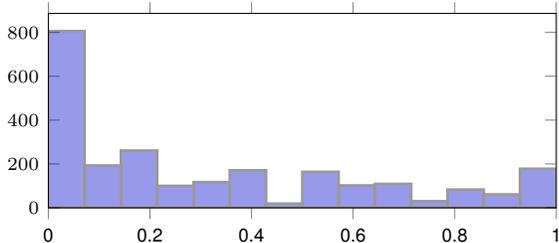
\begin{figure}[!t]
\centering
\begin{tikzpicture}
\begin{axis}[ybar,width=\linewidth,height=.5\linewidth,
	xmin=0,xmax=1,ymin=0,
	xtick={0,0.2,0.4,0.6,0.8,1}, xticklabels={0,0.2,0.4,0.6,0.8,1},
	]
\addplot[ybar interval,fill=blue!80!black,opacity=.4] coordinates
	{(0,806) (0.07142857,193) (0.14285714,261) (0.21428571,100) (0.28571429,117) (0.35714286,171) (0.42857143,19) (0.5,164) (0.57142857,102) (0.64285714,109) (0.71428571,30) (0.78571429,83) (0.85714286,61) (0.92857143,178) (1,0)};
\end{axis}
\end{tikzpicture}
    \caption{Cellularity score distribution.}\label{fig:distribution}
\end{figure}

In our experiments we used $2,395$ patches of $512\times512$ pixels in size, extracted from $96$ haematoxylin and eosin (H\&E) stained whole slide images (WSI) acquired from 64 patients. Each patch in the training set has been assigned a tumor cellularity score by an expert pathologist. In Figure~\ref{fig:distribution}, we present a distribution of the cellularity scores in the dataset.

Besides the image data, we used the annotations ($X$ and $Y$ coordinates) to identify lymphocytes, malignant epithelial, and normal epithelial cell nuclei in the additional $153$ patches. Using these weak annotations, we generated the segmentation masks that were used in our experiments. Here, at each $XY$ location, we simply fit a blob of $15$ pixels in diameter. In Figure~\ref{fig:nuclei} we present the generated masks for various classes.

\subsection{Weakly-supervised cell segmentation}

In this study, characteristic features of the data present a serious challenge for developing segmentation models: 1) the data features no segmented cells, only their coordinates (that is why we use semi-supervised segmentation); 2) annotated nuclei are present only in 154 microscopic images, each containing 0-50 malignant cells. However, cell segmentation is not a distinct goal of this study. As mentioned in Section \ref{introduction} and in~\cite{cellularity, rajan2004change, boucheron}, cellularity within the tumor area is assessed by estimating the percentage area of the overall tumor bed comprised of invasive tumor cells.  Aggregated area of individual invasive cell areas serves as a proxy and does not represent the ultimate cellularity value. Cellularity is affected by cell density, localization, and tissue structure.  We use segmented cells essentially as an interpretable visualization of an invasive tumor within the tumor bed.

In our ablation studies, we evaluated our model in four different settings to find how different design choices influence the segmentation accuracy and generalization. Namely, we compared the model as described in Section~\ref{sec:method} with a standard U-Net decoder against an FPN decoder, and with the encoder initialized randomly against the encoder initialized with weights pretrained on ImageNet. In all settings, the model was trained for 150 epochs with the Adam optimizer and gradually decreasing learning rate from $10^{-4}$ to $10^{-5}$.

To obtain training patches, we downscaled the microscopy images $\times2$ times, randomly cropped a $256\times256$ area, and rescaled pixel values from $[0, 255]$ to $[-1, 1]$. As mentioned previously, segmentation targets were generated as 4-channel masks with round blobs, 15 pixels in diameter (the characteristic nucleus size), drawn in the nuclei centers. During training, we dynamically augmented images with vertical and horizontal flips, rotation, gamma, hue, and saturation utilizing the \emph{Albumentations} library~\cite{albumentations}.

\begin{table}[ht!]
\centering
\small
\caption{Segmentation results: the Jaccard index for different decoders and initializations.}
\begin{tabular}{ccccc}
\toprule
\textbf{Initialization} & \textbf{Standard decoder} & \textbf{FPN decoder} \\ \midrule
Random & 0.35 & 0.47 \\
ImageNet & 0.50	& \textbf{0.53} \\
\bottomrule
\end{tabular}
\label{tab:segmentation}
\end{table}

\begin{table}[ht!]
\centering
\small
\caption{Cellularity MSE with 95\% confidence intervals for the segmentation-based (first row) and for the end-to-end methods. Our results demonstrate the importance of ImageNet pre-training. C in the parentheses indicates classification, R -- regression and S -- segmentation.}\label{tab:cellularity}
\begin{tabular}{lcc}
\toprule
\multirow{2}{*}{\textbf{Model}} & \multicolumn{2}{c}{\textbf{Initialization}} \\ \cmidrule(l){2-3} 
 & \textbf{Random} & \textbf{ImageNet} \\ \midrule
GBT & 0.023 [0.019-0.026] & 0.022 [0.019-0.026]\\
\midrule
Resnet34 (SR) & 0.013 [0.011-0.015] &  0.013 [0.011-0.015] \\\midrule
ResNet34 (R) & 0.015 [0.013-0.018] & 0.011 [0.010-0.012]\\
ResNet50 (R) & 0.025 [0.022-0.028] & 0.011 [0.009-0.012]\\
Xception (R) & 0.017 [0.015-0.020] & \textbf{0.010 [0.009-0.012]}\\
Xception (C) & \textbf{0.012 [0.010-0.014]} & \textbf{0.010 [0.009-0.012]}\\
\bottomrule
\end{tabular}
\end{table}

\begin{table}[ht!]
\centering
\small
\caption{Cellularity Kappa (4 class binning) and Intra-Class Correlation Coefficient (ICC) with 95\% confidence intervals for the segmentation-based ($1^{st}$ and $2^{nd}$ rows) and for the methods predicting cellularity directly, without segmentation. All the models here utilize ImageNet pre-training. C in the parentheses indicates classification, R -- regression and S -- segmentation.}
\label{tab:cellularity_icc_kappa}
\begin{tabular}{lcc}
\toprule
\multirow{2}{*}{\textbf{Model}} & \multicolumn{2}{c}{\textbf{Metric}} \\ \cmidrule(l){2-3} 
 & \textbf{Kappa} & \textbf{ICC} \\ \midrule
GBT & 0.571 [0.520-0.622] & 0.787 [0.744-0.823]\\\midrule
Resnet34 (SR) & 0.658 [0.604-0.704] & 0.865 [0.835-0.891]\\\midrule
ResNet34 (R) & 0.649 [0.599-0.700] & 0.868 [0.840-0.892]\\
ResNet50 (R) & 0.652 [0.603-0.701] & 0.867 [0.844-0.894]\\
Xception (R) & 0.669 [0.616-0.713] & 0.881 [0.853-0.904]\\
Xception (C) & \textbf{0.689 [0.642-0.734]} & \textbf{0.883 [0.858-0.905]}\\
\bottomrule
Peikari et al.~\cite{peikari2017automatic} & 0.38-0.42 & 0.75 [0.71-0.79]\\
Akbar et al.~\cite{akbar2019automated} & --- & 0.83 [0.79-0.86]\\
\bottomrule
\end{tabular}
\end{table}

\begin{figure*}[!t]
\centering
\subfloat[Image]{\includegraphics[width=0.22\linewidth]{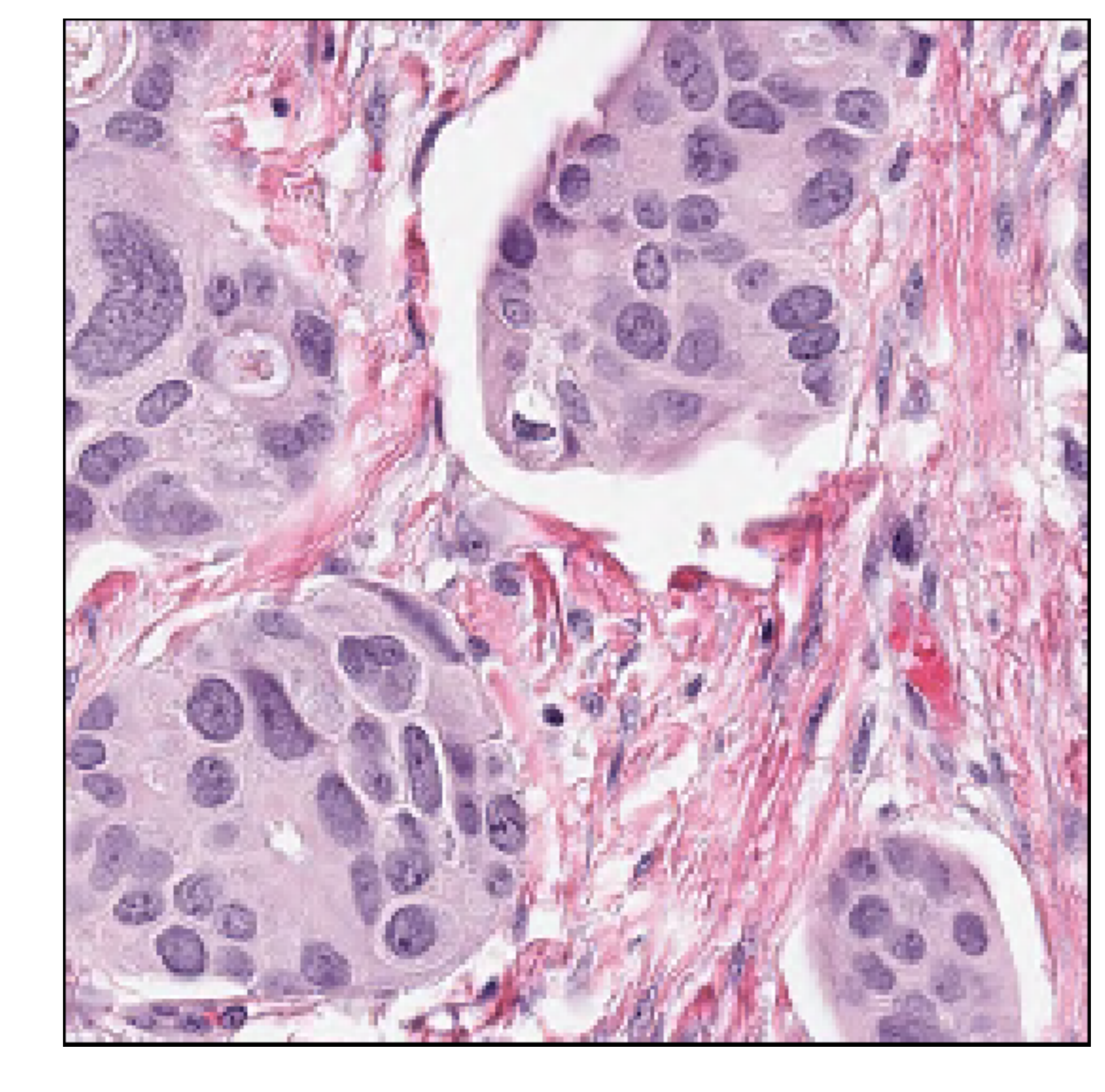}}
\hfill
\subfloat[Mask]{\includegraphics[width=0.22\linewidth]{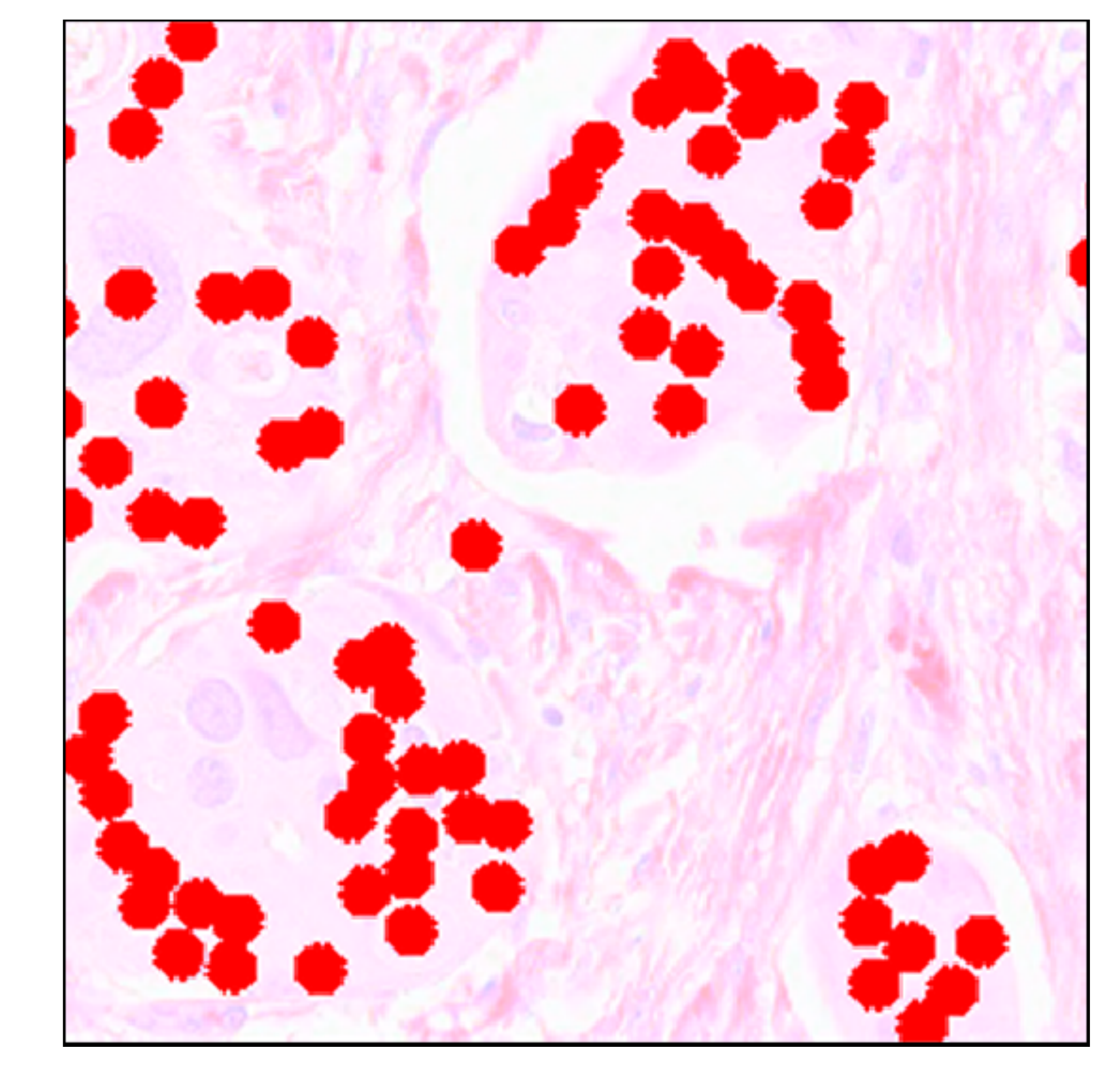}}
\hfill
\subfloat[Prediction]{\includegraphics[height=0.22\linewidth]{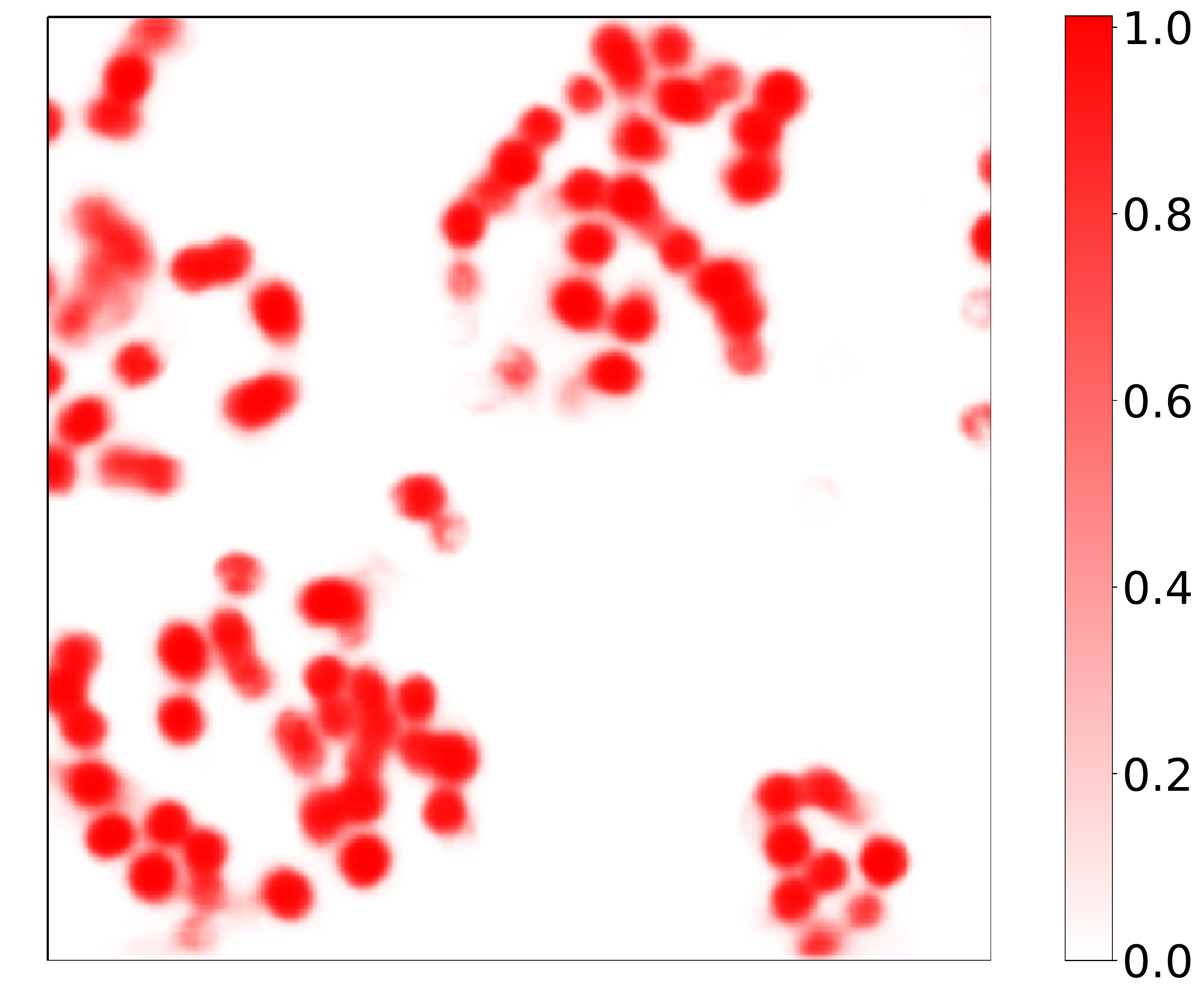}}
\hfill
\subfloat[LoG]{\includegraphics[width=0.22\linewidth]{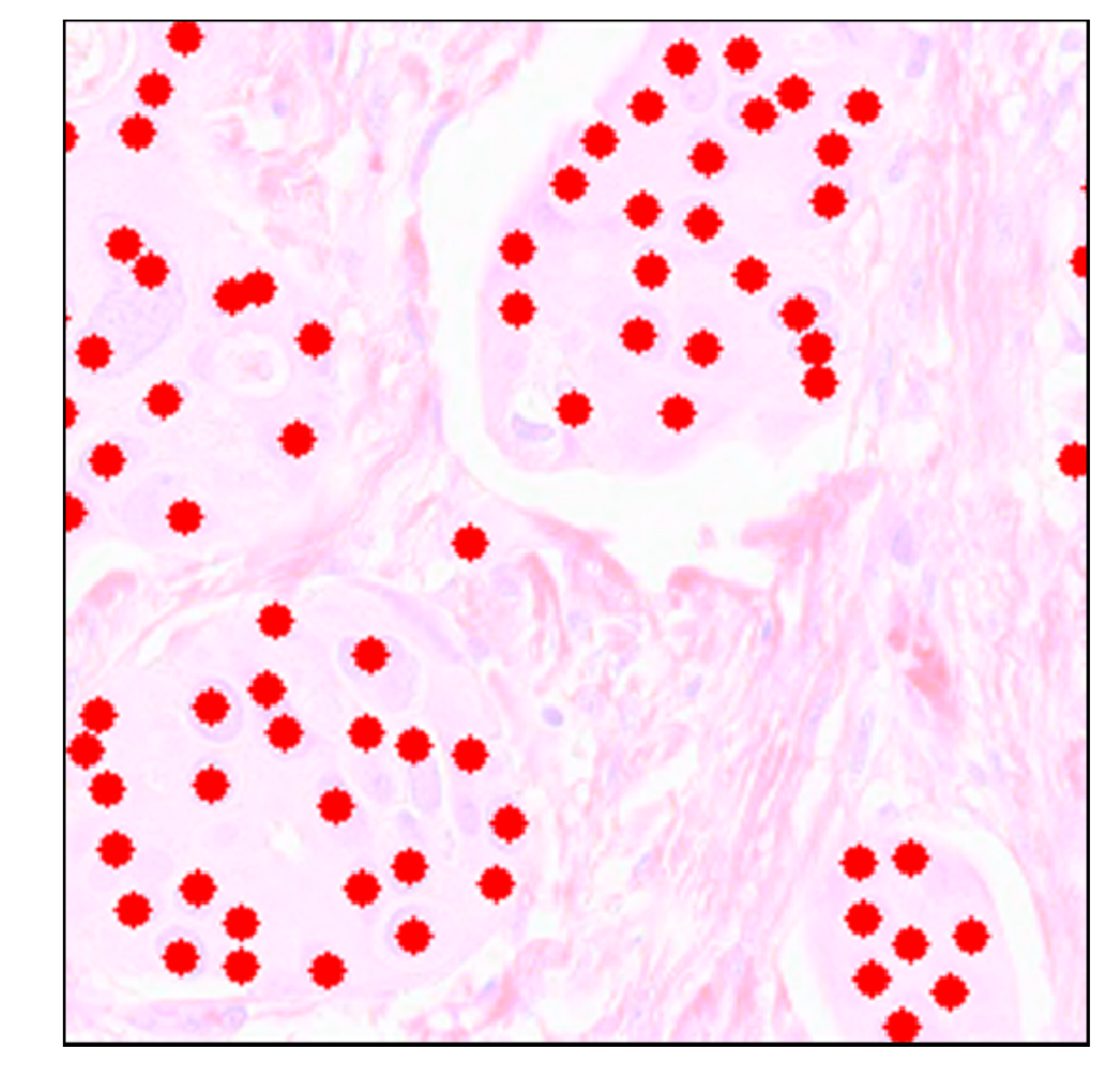}}

\caption{Examples of the generated segmentation masks in the \emph{Malignant} channel. Left to right: (a) original patch; (b) ground truth segmentation superimposed on the original image; (c) activation map; (d) nuclei blobs reconstructed from the activation map with the LoG method.}\label{fig:segmentation}
\end{figure*} 

In the first series of experiments, we evaluated segmentation quality as an important intermediate metric for the evaluation of our methods. The segmentation performance as a function of the decoder and initialization is shown in Table~\ref{tab:segmentation}. As we can see, the model with the feature pyramid decoder and encoder pretrained on ImageNet achieved significantly higher and more stable Jaccard index on the validation set than the alternatives. Figure~\ref{fig:segmentation} shows an example of generated segmentation masks in the Malignant channel and nuclei blobs reconstructed with the Laplacian of Gaussian method.

\begin{figure}[!t]
\centering

\begin{tikzpicture}
	\begin{axis}[scaled y ticks=false,xmin=0,xmax=104,ymax=0.06,ymin=0,width=\linewidth,height=.8\linewidth,xlabel={Training epochs},ylabel={Cellularity score MSE},legend style={at={(1.0,1.0)},anchor=north east},ytick={0,0.02,0.04,0.06,0.08,0.1}, yticklabels={0,0.02,0.04,0.06,0.08,0.1},
	y label style={at={(axis description cs:-0.1,.5)}}]
	\addplot[smooth,mark phase=1,mark=*,color=blue!50!black] table[col sep=comma,x index=0,y index=3] {mse.csv};
	\addplot[smooth,mark phase=7,mark=o,color=red!50!black] table[col sep=comma,x index=0,y index=4] {mse.csv};
	\addplot[smooth,mark phase=13,mark=oplus,color=green!50!black] table[col sep=comma,x index=0,y index=1] {mse.csv};
	\addplot[smooth,mark phase=19,mark=triangle*,color=magenta] table[col sep=comma,x index=0,y index=2] {mse.csv};
	\legend{{Training set, random init},{Test set, random init},{Training set, ImageNet init},{Test set, ImageNet init}}
	\end{axis}%
\end{tikzpicture}\vspace{-.2cm}
    \caption{Cellularity MSE evolution during training.}\label{fig:cellularity-performance}
\end{figure}
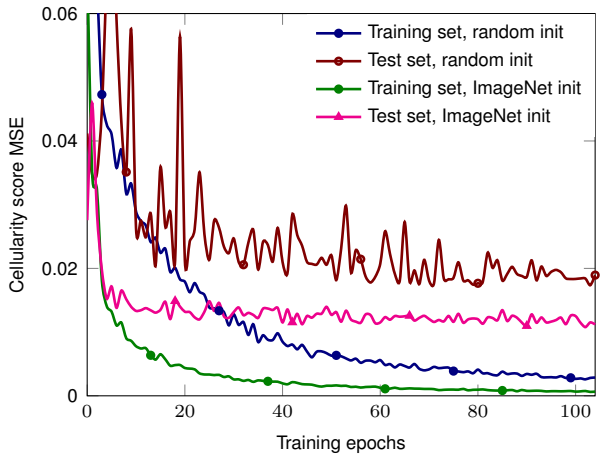 

\subsection{Segmentation-based cellularity assessment}

\paragraph{Prediction from the segmented cells.} As mentioned previously, we used the output of the segmentation model as input for the cellularity regressor and then trained this cascade end-to-end. We froze the segmentation model and stack its 4-channel output with a randomly initialized Resnet-34 in the regression setting. We trained the regression part with cellularity targets and MSE loss until convergence. Then we unfroze segmentation weights and fine-tuned both modules in an end-to-end fashion, as a single model. We repeated this experiment with Resnet-34 pretrained on ImageNet. In the latter case, we excluded the background channel from segmentation output to comply with the vanilla Resnet-34 architecture that has a 3-channel input.

In these experiments, we found that after fine-tuning the accuracy of segmentation itself slightly decreases, while the accuracy of the overall cellularity scoring increases. This is in line with~\cite{boucheron}, which found that perfect segmentation of nuclei figures does not ensure better classification of malignant objects from breast cancer tissues. This finding suggests that the two branches of future work, tumor bed segmentation and cellularity assessment, are relatively independent.

\paragraph{Feature extraction-based method.} In this series of experiments, we extracted the 81 features from segmentation masks as discussed in Section~\ref{sec:method} and trained the LightGBM~\cite{NIPS2017_6907} regression model with mean squared error (MSE) objective. The model was trained for 600 epochs with learning rate $0.01$. The maximum tree depth was set to 5; the number of leaves, to 8. These parameters have been selected through cross-validation.

\begin{figure}[t!]\centering
\pgfplotsset{every axis plot/.append style={
	line width=.6pt,
	mark size=1.75pt,
	mark repeat=10,
} }

\begin{tikzpicture}
\begin{axis}[
    xbar, xmode=log, xtick={10,100,1000,10000,100000}, xticklabels={10,100,1000,10000,100000},
    width=.9\linewidth, height=.8\linewidth, enlarge y limits=0.05, y=0.28cm, bar width=0.25cm,
    xlabel={Feature importance (logarithmic scale)},
    symbolic y coords={area@0.16,area@0.24,area@0.32,area@0.08,area@0.50,value@0.04,blb\_sum@0.08,blb\_sum@0.50,blb\_cnt@0.08,value@0.02,blb\_cnt@0.16,area@0.02,blb\_cnt@0.50,value@0.08,blb\_cnt@0.04,blb\_cnt@0.02,value@0.16,blb\_sum@0.04,blb\_cnt@0.24,blb\_sum@0.02,value@total,blb\_sum@0.16,area@0.04,blb\_sum@0.24,value@0.24,value@0.32,value@0.50},
    ytick=data,
    ]
    \addplot coordinates {(348.42,area@0.02) (382.61,blb\_cnt@0.50) (392.21,value@0.08) (506.51,blb\_cnt@0.04) (582.85,blb\_cnt@0.02) (654.68,value@0.16) (739.47,blb\_sum@0.04) (863.54,blb\_cnt@0.24) (940.65,blb\_sum@0.02) (1388.45,value@total) (1485.15,blb\_sum@0.16) (1646.8,area@0.04) (2833.58,blb\_sum@0.24) (4781.4,value@0.24) (8437.19,value@0.32) (182609.1,value@0.50)};
\end{axis}
\end{tikzpicture}\vspace{-.2cm}
\caption{GBT top feature importance.}\label{fig:feature-importance}\vspace{.5cm}
\end{figure}
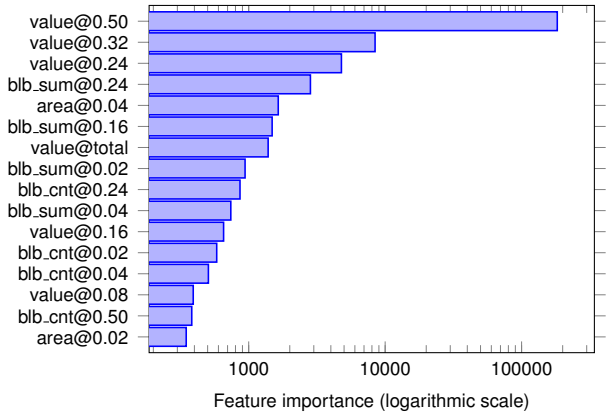

We report LightGBM accuracy in Table~\ref{tab:cellularity} and show the resulting feature importance on Fig.~\ref{fig:feature-importance}. Feature importance was calculated based on the total gain of the loss function from the splits formed according to this feature. As expected, all highly important features come from the \emph{Malignant} channel. The most important feature is the total activation above $0.5$ threshold, and the second and third most important features are the activations above $0.32$ and $0.24$ thresholds, as expected since activations at different thesholds are highly correlated, and the segmentation quality at threshold $0.5$ was the best, so the feature based on this mask is a natural candidate for the most important feature. Activations at lower thresholds provide additional value, but a big part of the information that they contain has already been conveyed via the $0.5$ threshold feature. Interestingly, malignant cell count (detected at threshold $0.24$) is only the $9$th feature in order of importance.

\subsection{Direct cellularity assessment from the raw images}

In our final experiments, we evaluated several deep neural architectures that take the original microscopy images as input and output the cellularity score without intermediate segmentation. Similarly to previous experiments, we trained the models with random \textit{He initialization}~\cite{he2015delving} or initialized them with weights pretrained on ImageNet. In all cases, ImageNet initialization was superior to random, and the overall accuracy was slightly better than for the models with intermediate segmentation. The Xception model implemented in a \emph{classification} setup with random initialization performed slightly better than its counterparts (MSE 0.012 vs. 0.017-0.025). Although small, this difference could possibly be attributed to the known regression-to-mean problem of the continuous regression with $L_2$ loss (e.g., see a well-explained example for a colourization application in~\cite{Zhang_2016}). The mean squared error of cellularity prediction as a function of the training epoch for different initializations is shown in Figure~\ref{fig:cellularity-performance}. All the performance evaluation metrics are presented in Table~\ref{tab:cellularity} and Table~\ref{tab:cellularity_icc_kappa}. 

\subsection{Discussion}

As we can see in Table~\ref{tab:cellularity} and Table~\ref{tab:cellularity_icc_kappa}, direct cellularity assessment method slightly outperforms the segmentation-based approach, where the regression module works on top of the segmentation feature extractor. We believe that performance improves due to two main reasons. First, segmentation models were not trained on accurate segmentation masks but rather on approximate masks generated from weakly supervised labels. Second, the cellularity score depends not only on the tumor masks but also on a broader set of features, some of which could be lost during the segmentation step.

While we note the record results of our end-to-end models, we believe that the modular form of the prediction pipeline provides benefits that more than compensate for this small difference in the final score.

The segmentation-based approach has two significant advantages: generalizability and interpretability. In practice, the data used for medical imaging tasks comes from different hospitals and is collected by different hardware. Images may differ in quality, level of noise, color and brightness distributions. In~\cite{iglovikov2017pediatric}, the authors proposed to use segmentation to clean and standardize the data, which helps with overall robustness and performance of various task-specific models.

Better interpretability is achieved by the fact that we can visually verify the quality of the intermediate step, i.e., segmented tumors. Furthermore, the decision trees model allows to estimate the feature importance for every feature based on the information gain. If segmented tumors are correct, and the most informative features make intuitive sense, we obtain additional confidence in our model, which is very important in the medical setting.

\section{Conclusion}\label{sec:conclusion}

In this paper, we evaluate three automatic methods to assess the cellulalarity of residual breast tumors in H\&E stained samples. Our first method leverages the weakly-supervised segmentation masks as inputs for deep CNN. We believe that this method will be more generalizable and robust towards the data acquisition and easier to interpret.

Our second method that leverages feature extraction from the weakly-supervised segmentation mask yields the highest score among the all previously published feature extraction-based methods \cite{akbar2019automated, peikari2017automatic}. 

Finally, the third method in this study is an end-to-end approach that predicts the cellularity score without any intermediate segmentation step. Although it is attractive and produces the best results it lacks interpretability of the segmentation-based methods and could perform best due to the dataset bias.

The main limitation of this study is the dataset size and the weak labels for the segmentation model. We think that given a bigger dataset and good quality annotations, segmentation-based approach could produce better results that less deviate from the end-to-end trained models.

\section*{Acknowledgements}

The work of Sergey Nikolenko was supported by the Russian Foundation for Basic Research grant no.~18-54-74005.
The authors thank the anonymous reviewer \#2 for the constructive suggestions that helped to improve the article.

{\small
\bibliographystyle{ieee}
\bibliography{references}
}
\end{document}